\documentclass[preprint]{revtex4-1}
\usepackage{graphicx}
\usepackage{dcolumn}
\usepackage{bm}
\usepackage{color}
\usepackage{soul}

\usepackage{amsmath}
\usepackage{amssymb}
\usepackage{natbib}

\newcommand{\nc}{\newcommand}
\nc{\ta}{\tilde{a}}
\nc{\I}{$I$}
\nc{\II}{$II$}
\nc{\III}{$III$}
\nc{\nn}{\nonumber}

\nc{\XYZ }{\bf} %
\nc{\ABC }{\st}

\begin{document}

\bibliographystyle{naturemag}

\title{
Thermal effects - an alternative mechanism for plasmonic-assisted photo-catalysis}

\author{Yonatan Dubi,$^{1,4}$ Ieng Wai Un,$^{2,3}$, Yonatan Sivan$^{3,4\ast}$
\\
\normalsize{$^{1}$Department of Chemistry, Ben-Gurion University, Israel }\\
\normalsize{$^{2}$Unit of Electro-Optics Engineering, Ben-Gurion University, Israel}\\
\normalsize{$^{3}$Joan and Irwin Jacobs TIX Institute, National Tsing Hua University, Taiwan}\\
\normalsize{$^{4}$ Ilse Katz Center for Nanoscale Science and Technology, Ben-Gurion University, Israel}\\
\normalsize{$^\ast$To whom correspondence should be addressed; E-mail: jdubi@bgu.ac.il}
}

\date{\today}

\begin{abstract}
  Recent experiments claimed that the enhancement of catalytic reaction rates occurs via the reduction of activation barriers driven by non-equilibrium (``hot'') electrons in plasmonic metal nanoparticles. These experiments place plasmonic photo-catalysis as a promising path for enhancing the efficiency of various chemical reactions. Here, we argue that what appears to be photo-catalysis is in fact thermo-catalysis, driven by the well-known plasmon-enhanced ability of illuminated metallic nanoparticles to serve as heat sources. Specifically, we point to some of the most important papers in the field, and show that a simple theory of illumination-induced heating can explain the extracted experimental data to remarkable agreement, with minimal to no fit parameters. We further show that any small temperature difference between the photocatalysis experiment and a control experiment performed under uniform external heating is effectively amplified by the exponential sensitivity of the reaction, and very likely to be interpreted incorrectly as ``hot'' electron effects.
\end{abstract}

\maketitle

\section{Introduction}
Many chemical reactions are catalyzed in the presence of metallic nanoparticles (NPs). The catalysis ensues via low activation energy pathways which become accessible only in the presence of the NPs~\cite{liu2018metal}. 
Typically, high-temperatures are used to further catalyze these reactions. However, besides being highly energy-consuming, and beside the associated shortened catalyst lifetimes~\cite{thermal_shortening_catalyst_lifetime}
, thermal activation is non-selective, leading to accompanying undesired reactions to take place and to loss of yield and efficiency, see~\cite{hot_e_review_Purdue} and references therein.

Recently, it was suggested to catalyze chemical reactions via photo-excitation of the electrons in the NPs. Presumably, this can happen via first excitation of localized surface plasmons in the metallic NP. As the plasmons decay, their energy is transferred to the ``hot'' carriers - non-thermal electrons and holes with excess high energy. It was claimed that these ``hot'' carriers couple to the reactants, and reduce further the activation energy of the favourable reaction pathways, as a function of their number density (hence, as a function of the incoming light intensity). Although 
this explanation is at odds with a conventional calculation based on the Fermi golden rule (see discussion in~\cite{Dubi-Sivan-Faraday}), this description became popular and formed the basis to the emerging field of plasmonic-assisted photo-catalysis, see e.g.,~\cite{plasmonic-chemistry-Baffou,plasmonic_photocatalysis_Clavero,hot_es_review_2015_Moskovits,hot_es_review_2015,Valentine_hot_e_review} for some recent reviews. 

\begin{figure}[h!]
    \centering
    \includegraphics[width=12.5truecm]{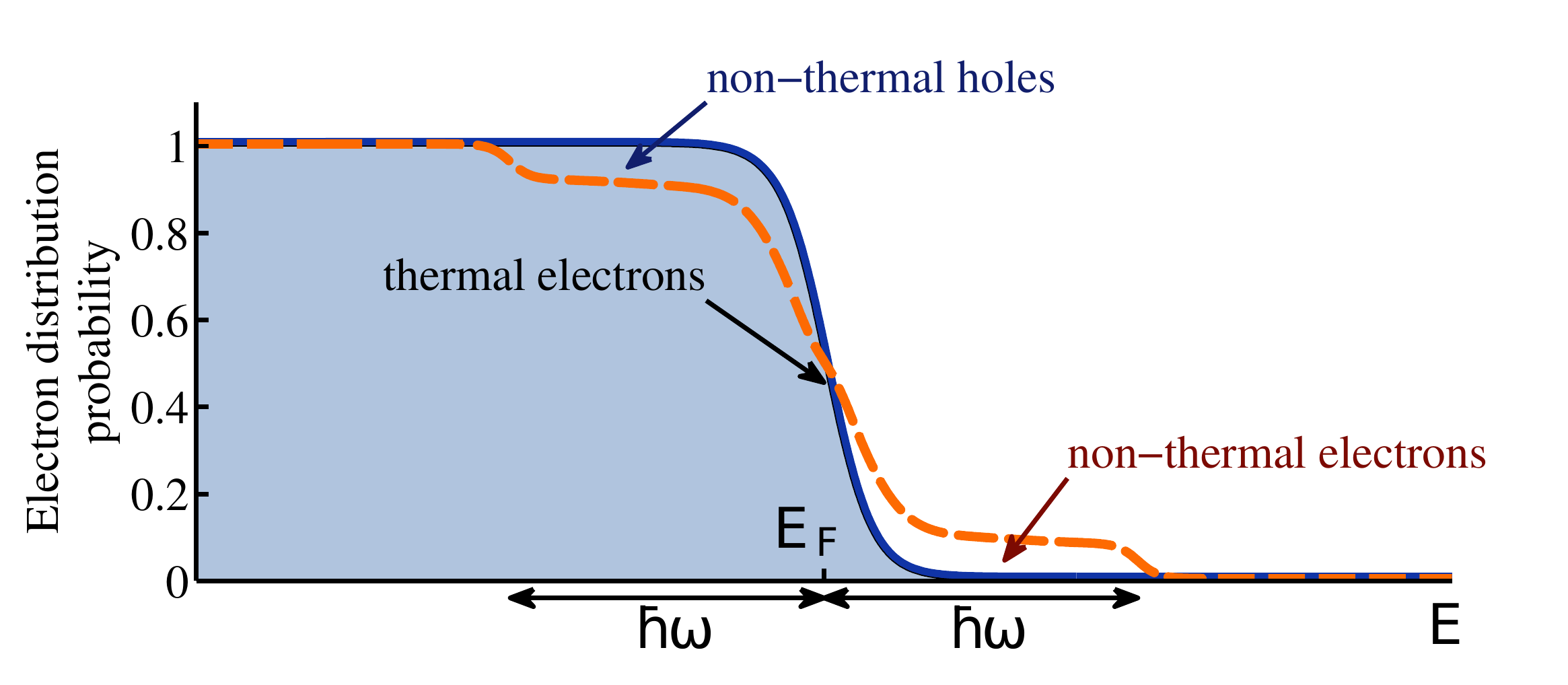}
    \caption{{\bf Schematic illustration of the electron distribution.} Blue solid line represents the equilibrium electron distribution in the absence of illumination. Orange dashed line represents the non-equilibrium electron distribution under illumination. It consists of thermal electrons near the Fermi energy which obey the Fermi-Dirac statistics, and non-thermal (the so-called ``hot'') electrons in two $\hbar \omega$-wide shoulders far from the Fermi energy which are not part of the Fermi-Dirac distribution. }
    \label{fig:hot_elec_sch}
\end{figure}

However, the relative importance of thermal and non-thermal effects remained an issue under debate. Specifically, the main question that arises in this context is how does the photon energy absorbed in the metal NPs split between the generation of high energy non-thermal (``hot'') electrons (i.e., those having energies far above the Fermi energy, and do not belong to the Fermi-Dirac distribution), and the regular heating of the NPs, which involves electrons close to the Fermi energy which do obey Fermi-Dirac statistics, see Fig.~\ref{fig:hot_elec_sch}. Remarkably, nearly all previous (experimental as well as theoretical) studies concluded that the thermal effects are negligible compared to non-thermal electron action, thus implying that the limitations associated with heating (discussed above) are circumvented. This conclusion led to a rapid growth of interest in plasmonic-assisted photocatalysis, mostly as a viable pathway towards cheap and efficient way to produce ``green'' fuels~\cite{plasmonic-chemistry-Baffou,plasmonic_photocatalysis_Clavero,hot_es_review_2015_Moskovits,hot_es_review_2015,Valentine_hot_e_review} that supposedly circumvents the known limitation of thermo-catalysis.

In this Article, we provide evidence that shows that, in contrast to the paradigm described above, the effects observed in some of the famous previous experiments are very likely due to mere heating of the NPs, such that non-thermal effects play a negligible role in plasmonic-assisted photo-catalysis. 
This pure thermal interpretation is based, initially (see Section~\ref{sec:theory}) on our first-principles theory in which the electron distribution and temperatures were computed self-consistently for the first time, see Ref.~\cite{Dubi-Sivan}. This theory showed that the power going to generation of ``hot'' electrons is an incredibly small fraction of the total absorbed energy, which thus goes in its entirety to heating. Then, we propose a purely thermal theory based on the Fermi golden rule and the Arrhenius Law which provides an alternative interpretation of the experimental data.

In Section~\ref{sec:experiments}, we focus on a few of the seminal papers on the topic, 
those which also provided (nearly) complete records of their experimental approach and data. First, we identify experimental errors that led the authors of these papers to underestimate the temperature rise, hence, the role of thermal effects. Second, we provide support to our claim by showing that the alternative theory described in Section~\ref{sec:theory} which only takes into account heating effects can not only explain experimental results in a simple and physically transparent way, but can also provide remarkable fits to the main results, with the minimum number of fit parameters. When possible, the values of these fit parameters are corroborated with a detailed calculation of the thermal response of the metal NP configurations used in the experiments. Finally, Section~\ref{sec:discussion} is devoted to a discussion of our results, of the severe limitations of the external heating control experiments employed so far and of possible future steps.


\section{Heating vs non-thermal effects: general argument}\label{sec:theory}
In a recent paper~\cite{Dubi-Sivan}, we developed a formalism to calculate the electron distribution in an illuminated NP, where the only physical assumption is that due to electron-electron interactions, the electron distribution relaxes towards a Fermi-distribution, a physically-intuitive assumption that underlies almost all previous theoretical studies of this problem. The main difference with respect to previous theoretical studies of this problem is that we ensured energy conservation in the electron-phonon-environment system; specifically, we accounted for the dependence of the heat transfer to the environment on the NP size and shape and on the environment's thermal properties. This approach allowed us to provide a correct quantitative prediction of the electron distribution, and to define and calculate an electron and phonon temperature unambiguously.

The main results of~\cite{Dubi-Sivan} were that (i) the electron and phonon temperatures are nearly equal and are  determined just by the illumination intensity, NP size and shape and the thermal conductance of the host, (ii) the efficiency of non-thermal (``hot'')-electron generation is $\sim 10^{-10} - 10^{-7}$ (for the low intensities typically used in photocatalysis experiments), i.e., only about one billionth of the energy provided by the illumination goes to creating non-thermal (``hot'') electrons, and the rest goes to heating. The latter result can be simply understood by noting that the electron relaxation time, which leads to thermalization, is about $10^6$ times faster than in standard gain materials (e.g., semiconductors or laser dye)~\cite{Sivan-OE-gain-2009,Dubi-Sivan-Faraday}. Accordingly, a $\sim 10^6$ stronger illumination intensity is required to balance it and to establish a substantial level of deviation from thermal equilibrium; these illumination levels are far above the damage threshold for metals, and the resulting temperatures are well above the melting temperatures. These claims are in agreement with the findings of Ref.~\cite{Petek_Nat_Phot_2017}, which showed experimentally that the number of high energy electrons that tunnel out from the metal to the surface is completely negligible in comparison to the number of high energy electrons directly generated in the dielectric (TiO$_2$) surface. They are also in accord with the findings in~\cite{Liu-Everitt-Nano-Letters-2019} which use what are some of the most careful temperature measurements in the context of plasmonic-assisted photocatalysis to show the absence of non-thermal effects.

The conventional way in which the temperature affects the rate of chemical reactions can be seen via the Arrhenius Law of chemical reactions. This Law, derived empirically in 1889 shows that the reaction rate $R$ is given by
\begin{eqnarray} \label{eq:arrn}
R &=& R_0 ~\mathrm{exp}\left(-\frac{E_a}{k_B T} \right),
\end{eqnarray}
where $k_B$ is the Boltzmann constant, $E_a$ is the reaction activation energy (to be more specific, the activation energy of the reaction's rate-limiting step), and $T$ is the temperature of the reactor; $R_0$ is a constant that depends on the details of the reactants (via the so-called collision theory), and if the reaction occurs primarily on the catalyst surface, then it also depends on details such as particle shape, density and number, the symmetry of its exposed facets, particle-molecule energy transfer rates, chemical interface damping etc., as well as measurement-dependent details such as sample degradation between different measurements.

In~\cite{Dubi-Sivan-Faraday}, we employed a Fermi golden rule type argument to show that under optical illumination, the reaction rate enhancement would be proportional to the number of ``hot'' electrons at the relevant energy, $R_0 \sim N_e$, which is in turn proportional to the illumination intensity, $I_{inc}$, thus yielding $R_0 \sim  I_{inc}$.  We emphasize that this simple theory is at odds with the claims on the dependence of the activation energy on the reaction rate, the same claims that underlie the growing interest in plasmonic-assisted photocatalysis.

This simple theory already shows that the faster reactions reported experimentally are extremely unlikely to originate from the presence of high energy non-thermal electrons. Indeed, although the absolute number of these ``hot'' electrons was calculated to be very small even under illumination~\cite{Dubi-Sivan}, the {\sl increase} in their number from dark to illumination is dramatic, up to 10-12 orders of magnitude (depending on the activation energy). This implies that the reaction rate should be faster by the same factor under illumination; clearly, this is much larger than the experimentally observed photo-catalysis.

An unavoidable conclusion is that the photo-catalytic rate enhancement is not due to high energy non-thermal (``hot'') electrons, but comes only from heating. Such a dependence arises from the dependence of the actual reactor temperature $T$ on the illumination intensity, which for sufficiently low intensity, can be written as
\begin{eqnarray} \label{eq:T_Iinc}
T(I_{inc}) &=& T_{dark} + a I_{inc},
\end{eqnarray}
where $T_{dark}$ is the temperature of the reactor when no illumination is present. The photothermal conversion coefficient $a$ depends on a number of system-specific parameters (NP size and shape, material, density and number, 
illumination wavelength, thermal properties of the host etc.)~\cite{Govorov_thermoplasmonics,thermo-plasmonics-multi_NP-Govorov,thermo-plasmonics-basics,thermo-plasmonics-review,Baffou_pulsed_heat_eq_with_Kapitza,thermo-plasmonics-multi_NP}. As shown below, $a$ can be calculated from first principles by summing properly the heat generated by all particles in the system. 
For higher intensities, the temperature (usually) grows more slowly (i.e., sublinearly) with intensity due to the increasing imaginary part of permittivity, resonance shifting and the resulting decreasing quality factor of the plasmonic cavity~\cite{SUSI} as well as due to the temperature dependence of the optical and thermal properties of the environment. For a thorough discussion of these effects, see Refs.~\cite{Sivan-Chu-high-T-nl-plasmonics,Gurwich-Sivan-CW-nlty-metal_NP} as well as the necessary temperature-dependent permittivity data presented in Refs.~\cite{Kippelen_JAP_2010,Shalaev_ellipsometry_silver,Shalaev_ellipsometry_gold,PT_Shen_ellipsometry_gold} and references therein.

Eq.~(2) implies that the dependence of reaction rate $R$ on temperature is a {\sl temperature-shifted Arrhenius Law}, i.e., a simple Arrhenius form, with a temperature that depends on the incident illumination intensity $I_{inc}$. In this context, it should be emphasized that although theoretical arguments were laid out in papers I-IV, 
they were not used to fit the data, and cannot be used for other, similar systems. Thus, Eqs.~(1)-(2) constitute the first ever attempt to quantitatively match experimental data of plasmonic-assisted photocatalysis experiment to any sort of theory. As we shall see, this attempt is extremely successful.

\section{Experimental data explained by heating}\label{sec:experiments}

To corroborate our claim regarding the dominance of thermal over non-thermal effects, we went back to some of the seminal papers in the field~\cite{Halas_dissociation_H2_TiO2,Halas_H2_dissociation_SiO2,plasmonic_photocatalysis_Linic,Halas_Science_2018} (denoted as papers I-IV hereafter) and extracted the experimentally measured data \footnote{This was done by digitizing the images, so some numerical errors $\sim 1\%$ might arise, but they do not affect our claims. }. Below, we point to the central shortcoming of each experiment, which led their authors to an extreme over-estimate of the non-thermal electron contribution to photo-catalysis, and show how their data can be fully understood and extremely well-fitted with the simple theory presented in Section~\ref{sec:theory}. For doing that, we had to distinguish between $T(I_{inc})$ (the actual temperature of the reactor), $T_{dark}$ (the temperature of the reactor in the dark) and $T_M$, which is the {\sl measured} temperature. Eq.~(2) can be then rewritten as
\begin{eqnarray} \label{eq:T_Iinc3}
T(I_{inc}) &=& T_{dark} + a I_{inc} = T_M + \tilde{a} I_{inc}.
\end{eqnarray}
As we describe below, in papers I-IV, $T_M$ is different from $T(I_{inc})$, an observation which explains their failure to distinguish thermal and non-thermal effects. When possible (see App.~\ref{app:T_calculation}), we compute the photothermal conversion coefficient $a$ by solving the heat equation for the relevant catalyst pellet geometries. It should be noted, however, that in papers I-IV, some of the heat generated by the absorption of light in the metal nanoparticles is removed by convection. Since a detailed calculation of the underlying equations for the relevant macroscopic structures is beyond the capabilities of standard computational approaches, we have computed the temperature due to heat diffusion under a set of reasonable assumptions, and estimated the convection to show that it would have only a small effect on the overall temperature profile, see App.~\ref{app:convection}).

\subsection{Analysis of papers I and II}\label{sec:analysis_I_II}

In Ref.~\cite[Paper I]{Halas_dissociation_H2_TiO2}, Mukherjee {\em et al.} demonstrated enhanced H$_2$ dissociation in the presence of illuminated Au NPs in a thick TiO$_2$ layer~\footnote{Note that the illustration in I shows that the Au NPs are dispersed just on the surface of the TiO$_2$ layer. However, judging from the reported mass concentrations etc., one has to conclude that the Au NPs are dispersed in the whole oxide thickness. }.

The central results in I are shown in their Fig.~2e where the reaction rate under illumination (in which case the measured temperature climbs to 30 C) is compared to the reaction rate in the dark, with the system being heated up externally to the same temperature. The observed $\sim 5.2$-fold increase in reaction rate under illumination 
was attributed to ``hot''-electron-induced catalysis due to an opening of a ``hot''-electron-initiated channel in the reaction energy surface, reducing the reaction energy barrier from $\sim 4.5$ eV to $\sim 1.7$ eV.

The entire analysis in I is based on controlling the reactor temperature. However, as was demonstrated in Refs.~\cite{Meunier_T_uniformity,Liu_thermal_vs_nonthermal,Liu-Everitt-Nano-Letters-2019} (and later acknowledged in IV), the temperatures can vary substantially inside the chemical reactor and they decay rapidly away from it; specifically, the temperature of the reactor can be very different (by 100s of degrees K) from the temperature measured by a thermocouple placed a few mm away. As shown below, indeed, the temperature measurements in I and II underestimated the reactor temperature, hence, led the authors to incorrect conclusions.


As an alternative explanation, we now show that the experimental data of I can be explained using a pure thermal effect, namely, Eqs.~(1)-(2). To start, from the reaction rate as a function of temperature in the dark (black circles in Fig.~4c of I, inset of Fig.~1(A)) we extract the reaction activation energy. Although the authors claim an energy scale of several electron volts, the experimentally measured value is $E_a\sim 0.23$eV. This is a surprisingly low value, which is not discussed in I. 
A possible explanation for such a low barrier is that the reaction is catalyzed by the oxide supporting the NPs via a heterolytic fragmentation path. Indeed, heterolytic cleavage reactions have been observed to have very low activation barriers~\cite{song2018heterolytic,joubert2006heterolytic}. 

Armed with this value for $E_a$ and the understanding that the real temperature of the catalytic surface is larger than the measured temperature under illumination, we ask a simple question: what temperature will give a rate which is $5.2$ times larger than the reaction rate measured in the dark? this is simple to answer, since all we need is to compare reaction rates given by Eq.~(1). The resulting temperature is $T \approx 362$K, an increase of $65$K compared to the ambient temperature $T_{dark} = 297$K, rather than just $6$K as assumed originally in I. From this, together with the known incident laser intensity $I_{inc} = 2.4 $W/cm$^2$, we extract the photothermal conversion coefficient (Eq.~(\ref{eq:T_Iinc3})), $\tilde{a} = 27.2$ K cm$^2$/W~\footnote{In contrast to papers III and IV described below, there was no sufficient information in I and II to enable a reliable calculation of the photothermal conversion coefficient $a$. }.

It is now a simple matter to understand the dependence of the reaction rates under illumination as a function of temperature. In Fig.~1(A) we plot the data from I; reaction rate as a function of temperature for different illumination intensities. The solid lines are the lines according to Eqs.~(1)-(3), with no fitting parameters (since all the information is already known). The temperature shifts as a function of intensities are plotted in the inset of Fig.~1(B), and the solid line is Eq.~(\ref{eq:T_Iinc3}) with $\ta = 27.2$ K cm$^2$/W. In Fig.~1(B) we plot the same data (rate as a function of temperature for different intensities), with the temperatures for each intensity shifted according to Eq.~(\ref{eq:T_Iinc3}). The resulting data falls onto a single exponential curve (Eq.~(1)). 
Thus, overall, the data from I shows excellent fit to a shifted Arrhenius Law with essentially no fitting parameters.

In Ref.~\cite[paper II]{Halas_H2_dissociation_SiO2}, the authors report a similar experiment (H$_2$ dissociation with Au NPs), the only essential difference from I is that the host is replaced, from TiO$_2$ (in I) to SiO$_2$. This results in a $\sim 150$-fold enhancement of the reaction rate under illumination (compared to $\sim 5$-fold enhancement in I under the same conditions). This result has a very simple, purely thermal explanation. The thermal conductance of SiO$_2$ is about $\sim 5-10$ times smaller than that of TiO$_2$. Accordingly, the temperature rise in the Au NPs on SiO$_2$ upon illumination 
is $\sim 5-10$ larger~\cite{thermo-plasmonics-basics,thermo-plasmonics-review}, so that the reaction rate (which depends exponentially on the inverse temperature, Eq.~(1)) becomes even more strongly enhanced, in fact, by a $25-100$-fold increase
, as observed experimentally.


\begin{widetext}

\begin{figure}[h!]
    \centering
    \includegraphics[width=16truecm]{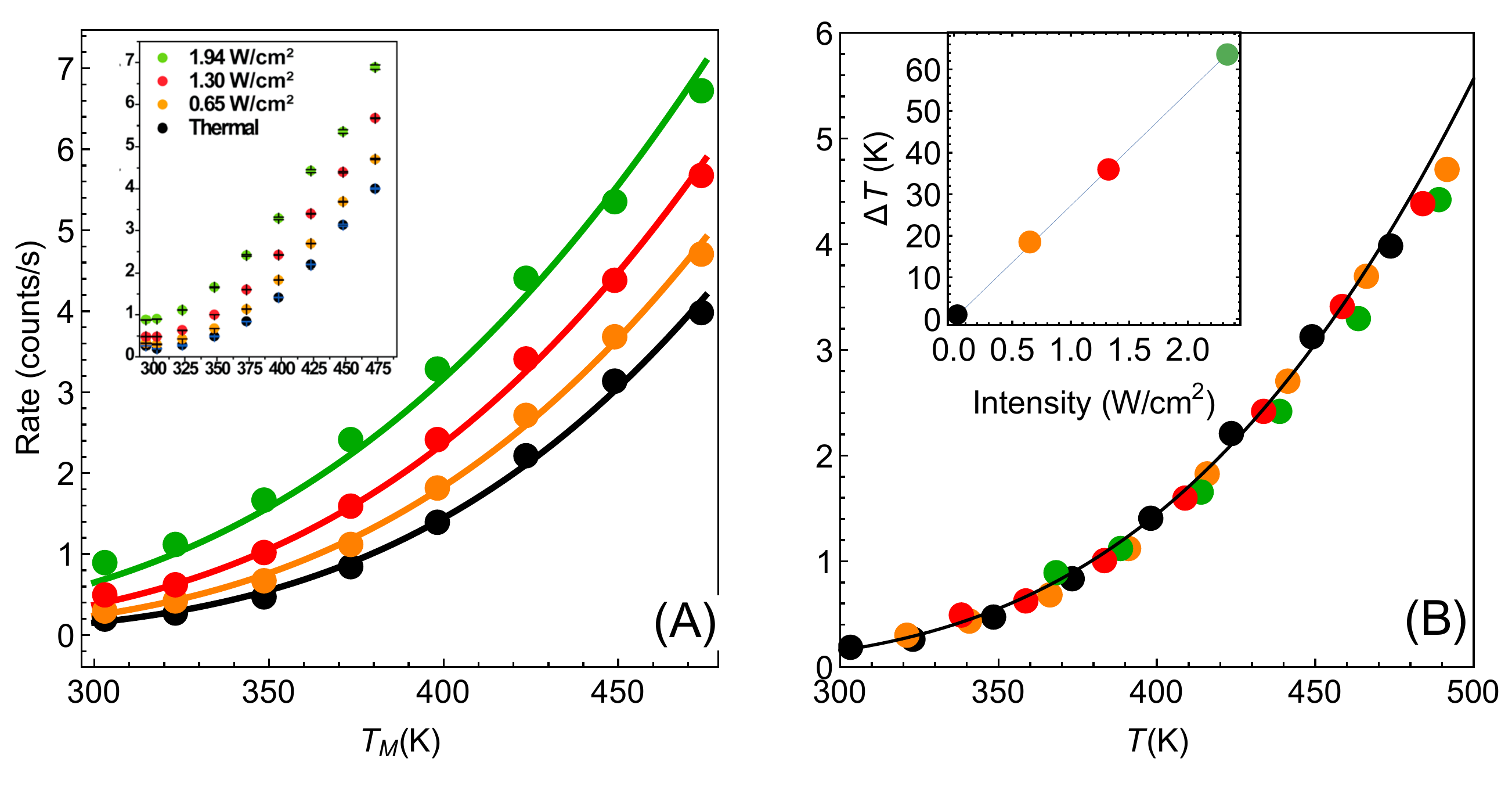}
    \caption{{\bf Temperature dependence of reaction rates (data from paper I)}. (A) Reaction rate as a function of (measured) temperature for different illumination intensities $I_{inc} = 0, 0.65, 1.3, 1.94$ W/cm$^2$ (black, orange, red and green points, respectively). Solid circles are data extracted from I (which is shown in its original form in the inset). Solid lines are fits to Eqs.~(1)-(\ref{eq:T_Iinc3}), with no fitting parameters, showing remarkable agreement between experiment and theory. (B) Same data as in (A), with the temperatures for each intensity shifted by the temperature rise given in Eq.~(\ref{eq:T_Iinc3}). With this shift, all data points fall on a single exponential curve ($R^2 = 0.9956$). Inset: temperature shifts as a function of intensity. 
    }
    \label{fig:Halas_III_I}
\end{figure}

\end{widetext}

\subsection{Analysis of paper III}\label{sec:analysis_III}

Another important example is the work of Christopher {\em et al.}~\cite[Paper III]{plasmonic_photocatalysis_Linic}, where the authors study O$_2$ dissociation in ethylene epoxidation. These authors placed $75$nm side-long Ag nano-cubes on $\alpha$-Al$_2$O$_3$ particles inside the reactor, and demonstrated that the reaction rate exhibits super-linear dependence on illumination intensity (Fig.~2a in III). Further, they demonstrated that upon illumination the reaction rate increases as a function of the external heating, manifested by an intensity-dependent reaction activation energy (Fig.~2c in III). Both these effects were attributed to plasmon-induced photocatalysis, and the former was specifically regarded as a unique characteristic of ``hot'' electron action, which cannot be observed by simply heating up the sample. The authors introduce an elaborate qualitative theory to explain these findings. They, however, dismissed the possibility of a thermal effect, based on a calculation they made in an earlier paper~\cite{christopher2011visible}, which we show below to be flawed. Nevertheless, once the reactor temperature is calculated correctly, the purely thermal model reproduces quantitatively the data of III with fantastic accuracy.

\begin{figure}[ht]
    \centering
    \includegraphics[width=8truecm]{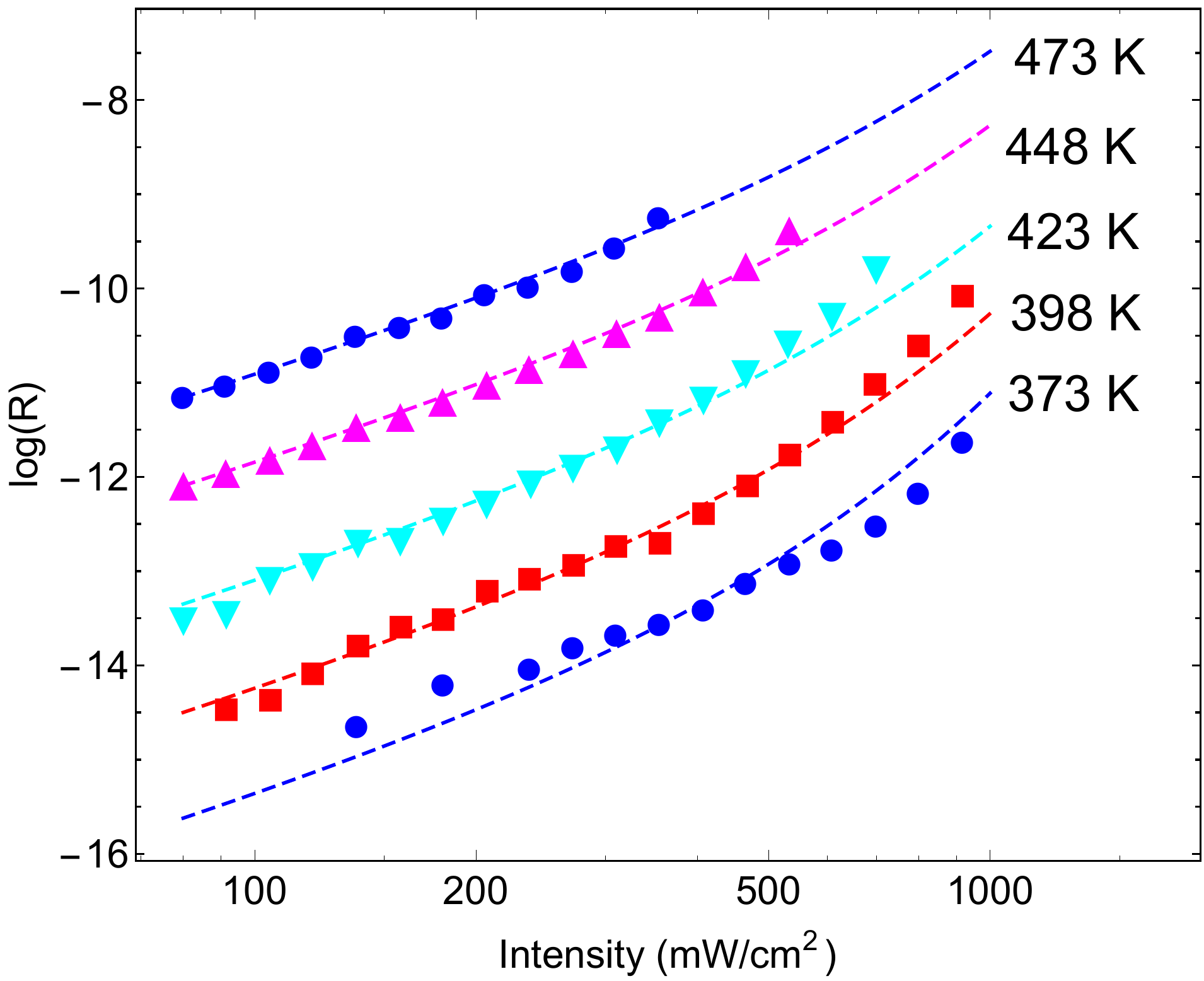}
    \caption{(Color online) {\bf Reaction rates at different (measured) temperatures as a function of incident intensity}. The symbols are data from paper III, whereas the dashed lines are theoretical curves based on Eqs.~(1)-(\ref{eq:T_Iinc3}).}
    \label{fig:Linic1}
\end{figure}

To show that, in Fig.~\ref{fig:Linic1} we plot the reaction rate as a function of illumination intensity for different (externally-measured) temperatures. 
From the measured data, one can extract the activation energy $E_a = 1.17$ eV and the photo-thermal conversion coefficient $\ta = 40$ K cm$^2$/W. It is important to note that 
$E_a$ and $\ta$ can be determined by any two data sets (say blue circles and red squares) and the other curves are then reproduced essentially without any additional parameters (except the pre-exponential coefficient $R_0$). Remarkably, the main acclaimed novelty in paper III, namely, the super-linear dependence of the reaction rate on illumination intensity is trivially reproduced by the temperature-shifted Arrenhuis Law, Eqs.~(1)-(2).

The fitted value of the photo-thermal conversion coefficient, $\ta = 40$ K cm$^2$/W, can be also obtained by an {\em independent} calculation under some reasonable assumptions based on a single particle temperature calculation~\cite{thermo-plasmonics-basics}, the procedure described in~\cite{thermo-plasmonics-multi_NP} for calculating the temperature rise due to the collective contributions of multiple NPs, and the sample description provided in the original manuscript~III itself. The details of this calculation are given in App.~\ref{app:Linic}. Notably, the value we obtained is much higher than in an earlier publication~\cite{christopher2011visible} on which the authors of III rely. However, the value obtained in Ref.~\cite{christopher2011visible} (namely, $a \sim 1.7 \times 10^{-5}$ K cm$^2$/W) was calculated for a {\sl single} NP, and did not take into account inter-NP heating (
see App.~\ref{app:Linic}), therefore underestimating the total heating by 6 orders of magnitude.

\begin{figure}[ht]
\centering
\includegraphics[width=8truecm]{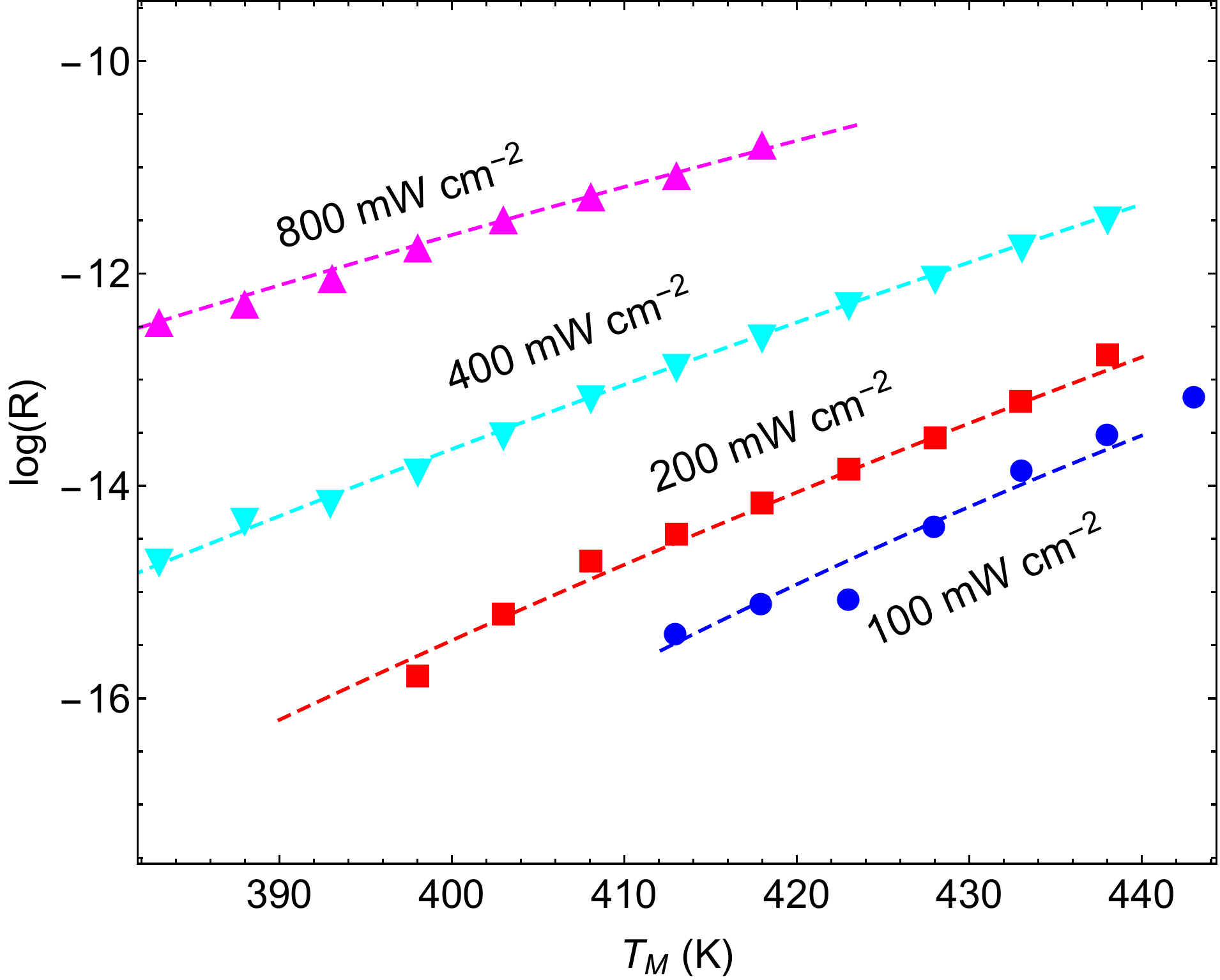}
\caption{(Color online) {\bf Reaction rates under different illumination intensity as a function of measured temperature}. The symbols are data from paper III, whereas the dashed lines are theoretical curves based on Eqs.~(1)-(3). Here, temperatures were varied externally using a heater.} \label{fig:Linic2}
\end{figure}

Further support for the thermal interpretation of III is provided in  Fig.~\ref{fig:Linic2}, where we show the reaction rate as a function of (externally-measured) temperatures for different illumination intensities. 
Using the same activation energy from the previous fit, only one data set is required to determine the photothermal conversion coefficient $a$, which is found to be $\ta \sim 160$ K cm$^2$/W; again, very good agreement is observed between the experimental data and the pure thermal explanation. This value is different from the value required for fitting the data of Fig.~\ref{fig:Linic1}, which may be due to the fact that different samples were used (this information is not available in paper III). 

Finally, we point out that the thermal theory presented here 
can reproduce also the Kinetic Isotope Effect reported in III (see App.~\ref{app:KIE_analy}
). Thus, essentially all the effects which were attributed to ``hot'' electrons in III can be fully reproduced with a thermal model that used the actual reactor temperature.

\subsection{Analysis of paper IV}\label{sec:analysis_IV}

\begin{figure}[ht]
    \centering
    \includegraphics[width=8truecm]{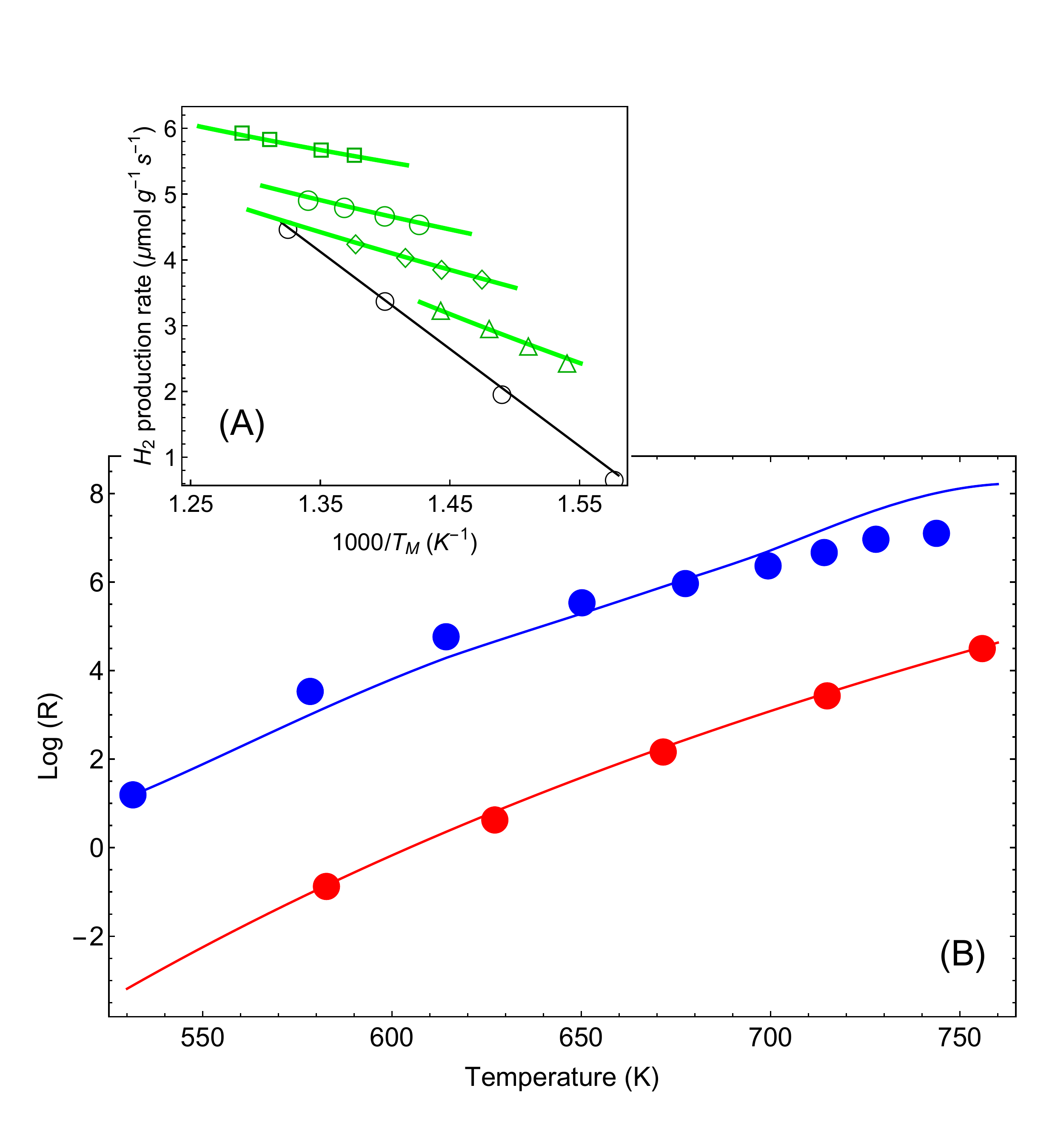}
    \caption{(Color online) {\bf Reaction rates under different illumination intensity as a function of inverse (average measured) temperature (data from paper IV)}. (A) Points correspond to the experimental data of Ref.~\cite{Halas_Science_2018} for the reaction rate for $\langle I \rangle = 0,1.6,2.4,3.2$ and $4$ W/cm$^2$ (empty circles, triangles, diamonds, disks and squares, respectively). The solid lines are a fit to Eqs.~(1)-(3). The parameters (activation energy $E_a$ and photothermal conversion factor $a$) are extracted from the open circles (in the dark) and the solid squares ($\langle I \rangle = 4$ W/cm$^2$) only. The curves for the rest of the data sets are obtained without additional fit parameters. Image borrowed from Ref.~\cite{anti-Halas_comment}. (B) Reaction rate as a function of (measured) temperature, in the dark (blue) and under illumination ($3.2, 4, 4.8, ...,9.6$ W cm$^{-2}$) with no external heating (red). Points are data from Ref.~\cite{Halas_Science_2018}, solid red line is an Arrhenius fit, and solid blue line is a shifted Arrhenius fit (Eqs.~(1)-(2)) with no additional parameters (except prefactor, see text). }
    \label{fig:IV_1}
\end{figure}

In a very recent paper~\cite[paper IV]{Halas_Science_2018}, the authors perform experiments which are similar to those presented in I and II, with several changes~\footnote{Similarly to I, the illustration in IV is limited to the immediate vicinity of the metal NP; this gives the impression that the Cu-Ru NPs are sparsely spaced on a solid oxide substrate. However, judging from the reported mass concentrations etc., one has to conclude that the metal NPs are dispersed in the whole oxide thickness, and that the oxide is highly porous (fill factor of 90\% air). }. First, the reaction considered is different (ammonia decomposition), meaning that the reaction activation energy would be different. More importantly, aiming at fixing the error in the temperature measurements of I-II, the authors used a thermal imaging camera to evaluate the temperature of the reactor. This is a crucial change, since by this they extract an (averaged) temperature value that allows them to approximately isolate the photo-thermal effects (page 12 in the supplementary material to IV). 
The authors then measure the reaction rate as a function of temperature for different illumination intensities (Fig.~\ref{fig:IV_1}(A)) and subtract the photothermal contribution. An Arrhenius fit to these data yields an {\em intensity-dependent} activation energy, which is the central result of paper IV. However, as shown in a recent Technical Comment~\cite{anti-Halas_comment}, the temperature measurements in IV suffer from systematic errors that invalidate its conclusions. Instead, we offer here again a pure thermal explanation based on Eqs.~(1)-(\ref{eq:T_Iinc3}) which remarkably reproduces the experimental data of IV.

For the sake of clarity, we briefly follow the procedure (described also above and in~\cite{anti-Halas_comment}). In Fig.~\ref{fig:IV_1}(A) we plot the reaction rate as a function of inverse measured temperature for different illumination intensities, taken from the data of IV. We fit a shifted Arrhenius Law to the data for the reaction rate in the dark, and under laser illumination of (average) intensity $3.2 $W/cm$^2$ and wavelength 550nm. These two data sets (empty circles and filled squares in Fig.~\ref{fig:IV_1}(A), respectively) yield $E_a\sim 1.3$ eV and $\ta = 180$ K cm$^2$/W. 
With these parameters we can fit the rest of the data, with no additional fit parameters. A remarkable agreement between the theory and the data is evident.

In similarity to the calculation performed for paper III (App.~\ref{app:Linic}), the fitted value of the photo-thermal conversion coefficient of paper IV can also be obtained by an {\em independent} calculation based on the sample description provided in the original manuscript IV itself as well as  the procedure described in Ref.~\cite{thermo-plasmonics-multi_NP}. However, since the source used in IV is pulsed, the expression for the temperature rise due to a single illuminated particle has to be based on the time-dependent solution, as described e.g., in Ref.~\cite{Baffou_pulsed_heat_eq_with_Kapitza}. Again, the procedure, described in App.~\ref{app:Halas_IV}, yields a value which is close to the one obtained from the fit.

Further support for the thermal interpretation of IV is provided in  Fig.~\ref{fig:IV_1}(B) where the reaction rate is plotted as a function of measured temperature. The data points, taken from paper IV, represent the following experimental procedure. The red points are the reaction rate in the dark (the temperature of the reactor is set by an external heater). The blue points, on the other hand, were obtained by illuminating the reactor with various intensities, measuring the resulting temperatures $T_M(I_{inc})$ (without any external heating), and plotting the reaction rate as a function of this temperature. The data shows an apparent increase of $\sim 2$ orders of magnitude in the reaction rate, one of the central results of IV.

To generate the shifted Arrhenius plot, we first fit the data in the dark to an Arrenhius curve (Eq.~(1), $E_a = 1.18$ eV), and then invert $T_M(I_{inc})$ to obtain the intensities $I_{inc}(T_M)$. The reaction rate, Eq.~(1), i.e., $R_0 \exp \left(-\frac{E_a}{T_M + \ta I_{inc}(T_M)}\right)$, is then plotted as a function of $T_M$, with $\ta = 180$ K cm$^2$/W and $E_a$ obtained from the previous fit, leaving only the prefactor $R_0$ as a fit parameter. The good fit to the experimental data demonstrates the consistency of our theory and confirms that the faster reaction under illumination is related to the fact that $T \gg T_M$; we expect the fit to improve once the thermo-optic response discussed above (\cite{Sivan-Chu-high-T-nl-plasmonics,Gurwich-Sivan-CW-nlty-metal_NP}) shall be included.

Notably, the temperatures our fit predicts are sufficiently high such that NP melting might be expected. This possibility is shown in App.~\ref{app:melting?} to have, at most, a mild effect on the reaction rate.



\begin{figure}[ht]
    \centering
    \includegraphics[width=8truecm]{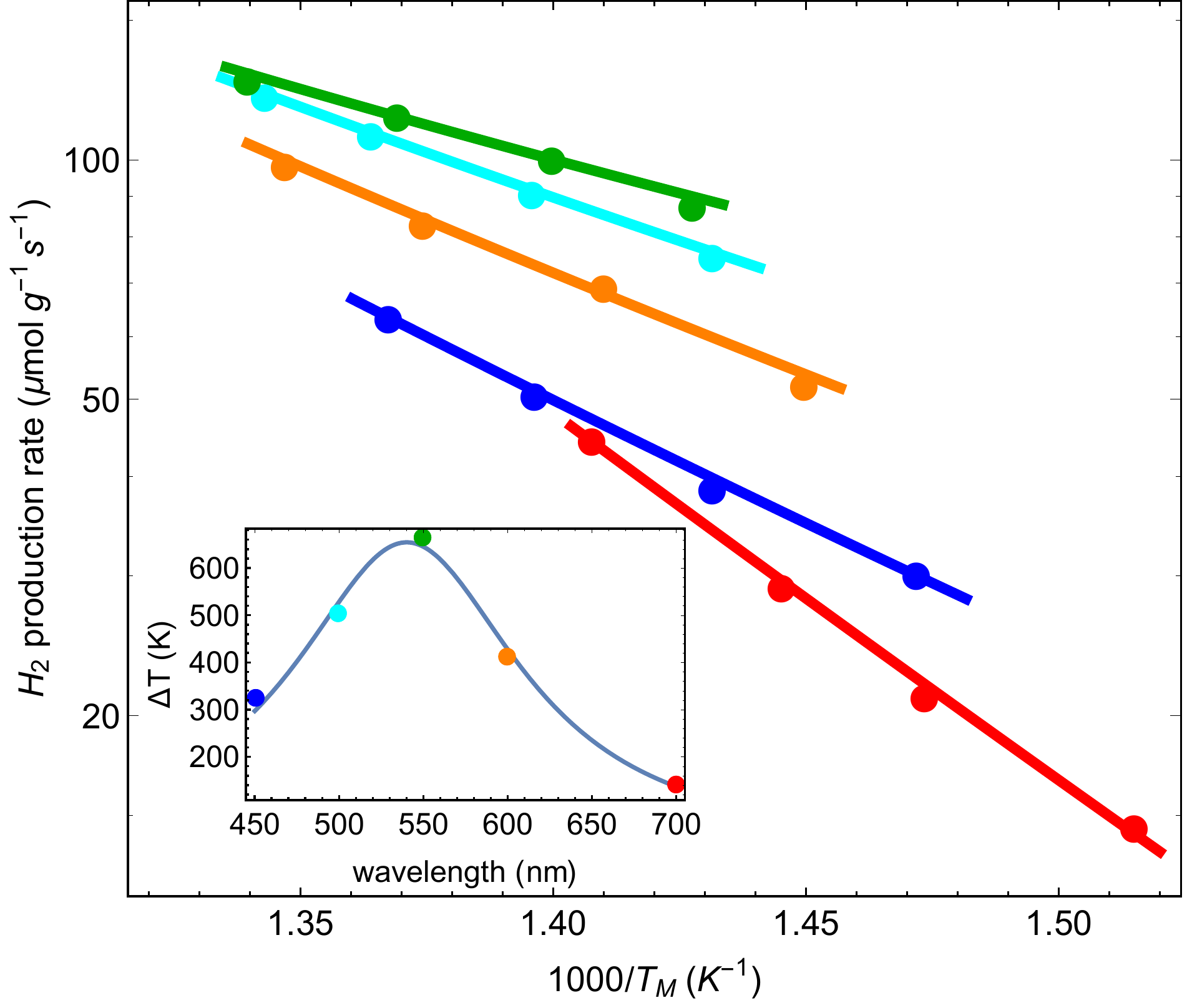}
    \caption{(Color online) {\bf Reaction rates as a function of inverse (average measured) temperature, for different illumination wavelengths (data from paper IV)}. Points correspond to original data of Ref.~\cite{Halas_Science_2018}, and lines are fits to a shifted Arrhenius (Eqs.~(1)-(3)). Inset: the fitted (effective) temperatures as a function of wavelength, showing maximal heating close to the plasmon resonance.
    \label{fig:IV_2}}
\end{figure}

Finally, we can follow the procedure used in Fig.~\ref{fig:IV_1}(A) for the data presented in IV regarding the dependence of the reaction rate on the laser wavelength. All we need to assume is that $a = a(\lambda)$ now depends on the wavelength, and hence $\Delta T = \Delta T(\lambda)$. In Fig.~\ref{fig:IV_2} we plot the reaction rate as a function of inverse temperature for different illumination wavelengths. The points are data from IV and the solid lines are fits to a shifted Arrhenius Law, Eqs.~(1)-(2). Again, we find excellent fit between our theory and the experimental data. 
The inset shows the resulting temperature rise $\Delta T = \ta I_{inc}$ 
(corresponding, roughly, to the maximal value reached in Fig.~2A of IV) as a function of wavelength, where the colored points correspond to the different curves in the main figure. The solid line is a fit to a Lorentzian, with a maximum corresponding to the plasmon resonance (at $540$ nm).

As an independent test, we computed the absorption cross-section of the Cu-Ru NPs using permittivity data from~\cite{optic_constant_noble_metals}; 
the resulting cross-section was essentially identical to those shown in the Supplementary of~IV (Fig. S12A). It is then a simple matter to fit the absorption cross-section to the data points. Notably, while the long wavelength side of the fit is satisfactory, the short wavelength side of the fitted curve exceeds the two extracted data points 
(data not shown). A similar discrepancy is seen in the deduced activation energy in IV (see their Fig. 2c); its origin might be partial conversion of absorbed electromagnetic energy into heat associated with interband transitions in Cu (at $\sim 2.1$eV), a possibility also raised in IV.

\section{Discussion}\label{sec:discussion}

The evidence we provided here suggests that in the specific papers discussed, there was nothing special in using plasmonic NPs for photo-catalysis; it proved to be yet another application for the use of plasmonic NPs as efficient localized heat sources~\cite{PT_imaging,Govorov_thermoplasmonics,thermo-plasmonics-basics,thermo-plasmonics-review,Halas-PT_treat,refractory_plasmonics,Halas-bubble1}. Specifically, our results demonstrate that the data of papers I-IV can essentially be explained with a simple Arrhenius theory, the only requirement is that the temperature of the reactor be evaluated accurately. In papers I-III, the origin of the discrepancy between the measured temperature and the actual reactor temperature is simple to understand; its origin is in the fact that a thermometer was placed away from the reactor pellet, thus discarding any temperature gradients which appear in the reactor and beyond it (as was recently discussed in Refs.~\cite{Meunier_T_uniformity,Liu_thermal_vs_nonthermal,Liu-Everitt-Nano-Letters-2019}).

In paper IV, the authors made substantial effort to overcome this, by using a thermal camera. However, even with this improvement, there may be several sources for temperature ambiguity. For instance, the use of a thermal camera for materials of low emissivity may result in a systematic temperature under-estimation~\cite{anti-Halas_comment}. Another possible source of error are temperature gradients within the sample due to the directional illumination, or even temperature transients which are nearly impossible to reproduce in a control experiment based on external heating (thermocatalysis). In fact, in~\cite{Liu-Everitt-Nano-Letters-2019} it was shown that temperature gradients should be expected even in the dark control experiments, either due to the gas flow or due to anisotropic external heating. Thus, any attempt to subtract the thermocatalysis control results necessarily leads to an incorrect interpretation of the difference between the thermal contributions in the dark control and the photocatalysis experiments as ``hot'' electron action. Similar difficulties will arise if one attempts to compute the temperature of the reaction - any small error will be incorrectly interpreted as ``hot'' electron action.

Since the temperature recorded by the camera is an average over space (and time), while the reaction rate is exponentially sensitive to temperature changes, this methodology effectively overlooks the fact that the reaction occurs preferably in the higher temperature regions (or times in which the temperature peaks), thus, necessarily underestimating the thermal contribution. This is nothing but the trivial mathematical statement that $\overline{R(T(\vec{r}))} \neq R(\overline{T(\vec{r})})$ (especially correct for exponential functions). The associated errors 
can in fact be huge. For example, for the conditions of IV, where the temperature drops to less than $50\%$ along the axis of the sample (see Fig.~\ref{fig:temp_uniformity_mti_pulse}(b)), there is an orders of magnitude difference between the reaction rates on the top and bottom of the pellet. This suggests that the inhomogeneities must be minimized in order to allow the distinction between thermal and non-thermal effects. This can be achieved by using thinner pellets, or ultimately, by studying a single particle~\cite{Cortes_Nat_Comm_2017,Gross_Nature_2017,Florian_Giessen_Hydrogenation,Vadai_Dionne_hydrogenation}.

The bottom line of the above discussion is that a thermocatalysis control experiments (i.e., using external heating to achieve uniform and steady-state heating of the pellet) can only identify ``hot'' electron contributions which are far larger compared to the errors associated with the temperature non-uniformities, transients and measurement accuracies; this can be tested e.g., by varying the NP density or pellet size. Since, as explained above, we expect the ``hot'' electron contribution to be very small~\cite{Dubi-Sivan,Dubi-Sivan-Faraday}, this means that a thermocatalysis control has essentially no chance to yield more than a non-tight upper limit estimate to the effect of the electron non-equilibrium on the reaction. 



An orthogonal approach for separating thermal from non-thermal effects is to perform the same measurements with illumination at gradually longer wavelengths. ``Hot'' electrons created under such low-energy illumination will not have enough energy to contribute to the reaction. Therefore, if the reaction is indeed based on a ``hot'' electron mechanism, a significant drop in the reaction rate will occur for sufficiently long wavelength
. This is, in fact, the principle underlying the use of ``hot'' electrons for photo-detection - a photon is detected only if it has sufficient energy to cross the Schottky barrier and travel to the detector on the semiconductor side; otherwise, the contact is considered to be Ohmic, see e.g., Refs.~\cite{Uriel_Schottky,Uriel_Schottky2,Moskovits_hot_es,Fabrizio_hot_es,Valentine_hot_e_review,Satoshi_TiN_hot_es}. A similar mechanism ensures ``hot'' electron action in upconversion experiments~\cite{Guru_APL_up_conversion,Guru_APL_up_conversion_exp}. In contrast, the thermal mechanism we propose predicts that under these conditions there will be no drop in the photo-catalytic enhancement, since the system will heat up even under low-energy illumination. Notably, wavelength dependence of the reaction rate is frequently recorded in plasmonic-assisted photocatalysis studies. We are not aware of any report of a sharp decrease of reaction rate for long wavelengths; this further supports our purely thermal interpretation of experimental data. Yet, the failure to observe such a sharp drop might be caused by the use of white light sources rather than monochromatic sources. Thus, more careful wavelength dependence studies might be worthwhile performing.

Having said all the above, it is important to mention that some previous papers reported photocatalytic action that cannot be explained just using thermal effects, e.g., the reaction selectivity reported in Ref.~\cite{Liu_Everitt_selectivity_Nat_Comm}. 
The theoretical approach of Ref.~\cite{Dubi-Sivan}, together with the detailed thermal calculations of Refs.~\cite{Baffou_pulsed_heat_eq_with_Kapitza,thermo-plasmonics-multi_NP,thermo-plasmonics-basics} (as demonstrated in the current manuscript), existing theory of electron tunnelling (see e.g., Ref.~\cite{Uriel_Schottky_2018}) and the vast knowledge accumulated on heterogeneous catalysis on the various chemical parameters that affect the reaction rate can now provide, for the first time, the necessary framework to analyze the relative efficiency of non-thermal and thermal effects and distinguish between optical and chemical aspects in these previously published papers, as well as in future papers on the topic.








\bigskip

{\bf Acknowledgements} The authors are grateful to Dr. A. Milo, Dr. Josh Baraban and Prof. M. Bar Sadan for valuable discussions and critical reading of the manuscript. YS was partially supported by Israel Science Foundation (ISF) grant no. 899/16. 

\bigskip

{\bf Author Contribution} YS and YD initiated the study, YD performed the thermal analysis, IWU performed the detailed temperature calculations. All authors contributed to the interpretation of the experimental data and writing the paper.


\appendix

\section{Detailed Temperature calculations}\label{app:T_calculation}
In this Supplementary Information Section, we compute the temperature of the catalyst pellet described in~\cite[paper III]{plasmonic_photocatalysis_Linic} and~\cite[paper IV]{Halas_Science_2018}, respectively. We describe in detail the assumptions employed in the calculations, the calculation procedure itself, and discuss the sensitivity of the results to the uncertainty in the various parameters.

\subsection{Detailed temperature calculations for paper III}\label{app:Linic}
In~\cite[paper III]{plasmonic_photocatalysis_Linic}, the catalyst pellet consisted of Ag nanocubes (edge length 75 nm) mixed with larger Al$_2$O$_3$ particles. For simplicity, we approximate the system as a (periodic) Ag NP array immersed in a uniform host material. Specifically, the nanocubes are approximated by nanospheres of radius $R = 55$ nm such that the volumes of the cubes and spheres are approximately the same. Based on the reported concentration of Ag (20 wt\%) in the composite~\cite[paper III]{plasmonic_photocatalysis_Linic} and the reported size of the catalyst layer~\cite{christopher2011visible} (thickness $H = $ 0.5 mm)~\footnote{The authors did not mention the sample area, therefore, we assume that it is uniform and similar in size to that of the quartz window of the reaction chamber.}, we can estimate that the average inter-particle separation is $p \approx 354$ nm.

For the host material, we set the (effective) permittivity to be $\varepsilon_{\textrm{h}} = (1 - f_{\textrm{v}})\varepsilon_{\textrm{air}} + f_{\textrm{v}}\varepsilon_{{\textrm{Al}_2\textrm{O}_3}}$ and the thermal conductivity to be~\cite{maxwell1881treatise,birdtransport,Pietrak-eff-kappa-compos-2014}
\begin{align}\label{eq:eff_kapa_h_linic}
\kappa_h = \kappa_{\textrm{air}}+\dfrac{3f_{\textrm{v}}\kappa_{\textrm{air}}}{\frac{\kappa_{\textrm{Al}_2\textrm{O}_3} + 2\kappa_{\textrm{air}}}{\kappa_{\textrm{Al}_2\textrm{O}_3} - \kappa_{\textrm{air}}} - f_v},
\end{align}
where $f_{\textrm{v}}$ is the volume fraction of oxide in the composite. The relatively more advanced effective medium formula used for the thermal conductivity is required due to the large differences between the conductivities of the constituents.

\begin{figure}[h]
\centering
\includegraphics[width=7truecm]{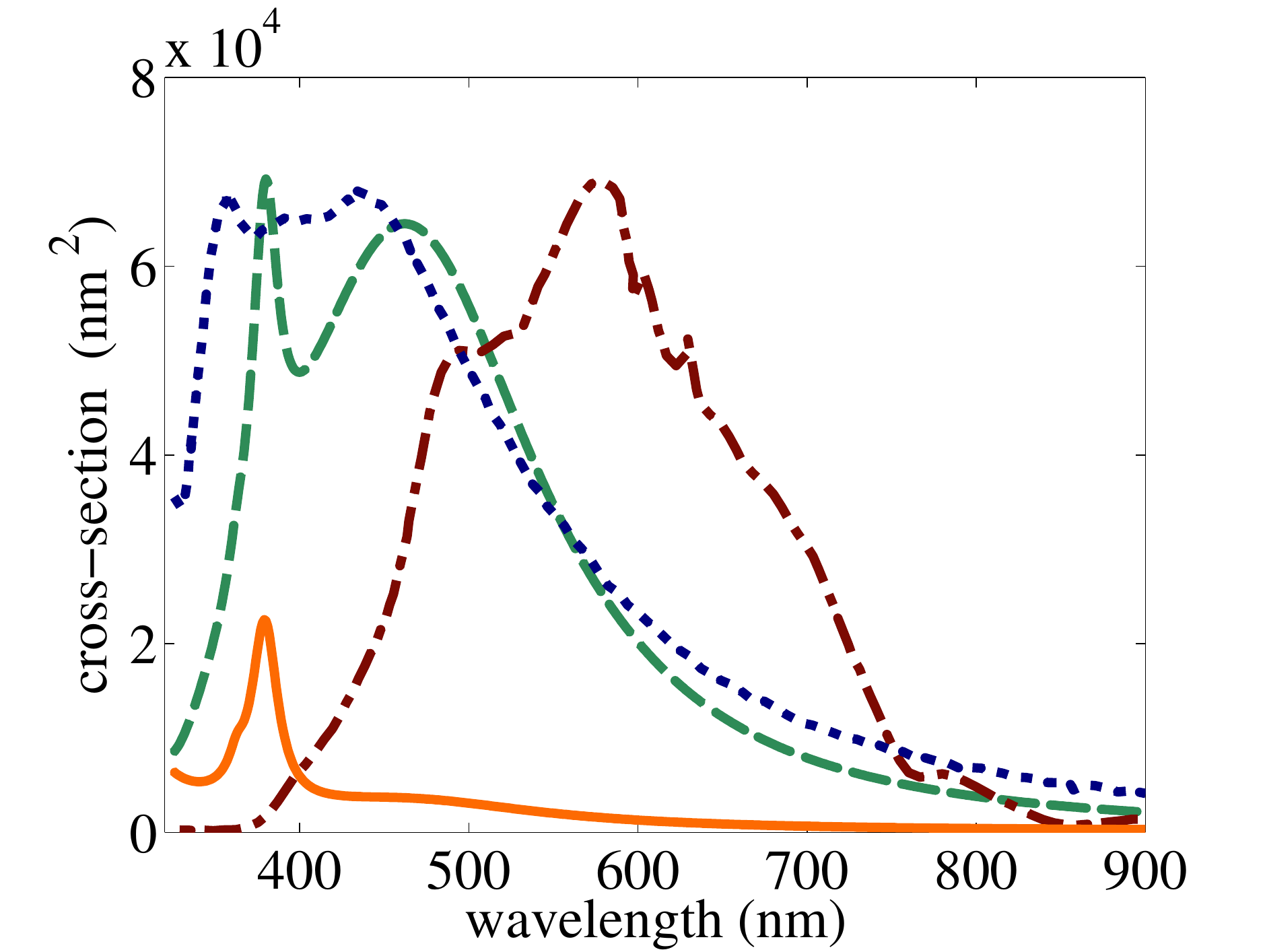}
\caption{(Color online) Calculated extinction (green dashed line) and absorption (orange solid line) cross-section of an Ag nanosphere of 55 nm in radius at temperature 373 K using the high temperature ellipsometry data of~\cite{Shalaev_ellipsometry_silver}. The measured extinction spectrum of Ag nanocube (blue dotted line) and the emission spectrum of the visible light source (red dashed-dotted line) $i_{\textrm{inc}}(\omega)$ are also shown.}\label{fig:ext_abs_crx_sec_ag_nsph}
\end{figure}

The settings for the nanosphere radius and the fraction of $\varepsilon_{\textrm{Al}_2\textrm{O}_3}$ in $\varepsilon_h$ are verified by comparing the calculated extinction cross-section with the measured extinction spectrum~\cite[paper III]{plasmonic_photocatalysis_Linic}, as shown in Fig.~\ref{fig:ext_abs_crx_sec_ag_nsph}. One can see that the differences in the extinction between the calculation and measurement are mostly in the region which the light source has a low spectral density.

The sample is subject to white light continuous wave (CW) illumination. The spectrum of the CW light source (copied from reference~\cite[paper III]{plasmonic_photocatalysis_Linic}) $i_{\textrm{inc}}(\omega)$ is shown in Fig.~\ref{fig:ext_abs_crx_sec_ag_nsph} and the incident intensity is $I_{\textrm{inc}} = \int i_{\textrm{inc}}(\omega) d\omega$ with a spot size of $A \sim 1$ cm$^2$ ($\sim 5.6$ mm in radius) which is assumed to be similar to the area of the quartz window of the reaction chamber.

The temperature distribution in the catalyst pellet $T({\bf r})$ can be obtained by solving the heat equation
\begin{align}\label{eq:general_heat_eq_steady}
\begin{cases}\kappa_m \nabla^2T({\bf r}) = -p_{\textrm{abs}}({\bf r}), & {\textrm{for }}{\bf r}\textrm{ in NPs,} \\
\kappa_h \nabla^2T({\bf r}) = 0, & {\textrm{for }}{\bf r}\textrm{ in host,}
\end{cases}
\end{align}
with appropriate boundary conditions at the surface of each NP~\cite{thermo-plasmonics-basics}. Here, $p_{\textrm{abs}}({\bf r})$ is the absorbed power density; it is related to the total (local) electric field ${\bf E}(\omega,{\bf r})$ via $p_{\textrm{abs}}({\bf r}) = \int \frac{\omega}{2} \varepsilon_m^{\prime\prime}(\omega,{\bf r}) |{\bf E}(\omega,{\bf r})|^2 d\omega$~\cite{Jackson-book}. At room temperature, the total (local) electric field and, thus, the absorbed power density can be obtained just by solving the Maxwell's equations. However, due to the large domain size and the huge number ($10^{12}$) of NPs, such numerical calculation could be time-consuming or even unfeasible.

To simplify the problem, we neglect the temperature dependence of the permittivities and the thermal conductivities; this will be justified a-posteriori by the modest temperature rise of only a few tens of degrees that shall be retrieved. Further, since $R \ll p < \lambda$, i.e., since the particle density is relatively low, we can apply the effective medium approximation such that we can write the incident field intensity as $i({\bf r},\omega) = i_{\textrm{inc}}(\omega) \exp(-z/\delta_{\textrm{skin}}(\omega)$ due to the absorption by the NPs, where $1/\delta_{\textrm{skin}}$ is the absorption coefficient experienced by the incident beam; the transverse profile of the illumination is assumed to be uniform. 

The absorption coefficient can be obtained by considering the change of spectral intensity per unit length along the propagation direction, namely,
\begin{align}
\dfrac{\Delta i(\omega,z)}{\Delta z} = -\dfrac{i(\omega,z) \sigma_{\textrm{abs}}(\omega) A/p^2}{A} \dfrac{1}{p} = -i(\omega,z)\dfrac{\sigma_{\textrm{abs}}(\omega)}{p^3},\nonumber
\end{align}
where $\Delta z \sim p$ is the thickness of one layer, $i(\omega,z)\sigma_{\textrm{abs}}(\omega)$ is the absorbed power per unit frequency per NP and $A/p^2$ is the number of NPs per layer, so $i(\omega,z) \sigma_{\textrm{abs}}(\omega) A/p^2$ is the total absorbed power per unit frequency; we divide it by $A$ to obtain the intensity loss per unit frequency. Then, the penetration (or, skin) depth of light to the sample is
\begin{align}\label{eq:skin_depth}
\delta_{\textrm{skin}}(\omega) \approx p^3/\sigma_{\textrm{abs}}(\omega).
\end{align}
For visible wavelengths, the skin depth ranges from 15 $\mu$m to 100 $\mu$m.

Since $\kappa_m \gg \kappa_h$, the temperature is uniform within each NP even if $p_{\textrm{abs}}({\bf r})$ is highly non-uniform~\cite{thermo-plasmonics-basics}. This further allows us to replace the spatial-dependent $p_{\textrm{abs}}({\bf r})$ in each NP by its spatial average over the NP at ${\bf r}_i$, namely,
\begin{align}\label{eq:avg_pabs_i}
\bar{p}_{\textrm{abs},i} &= \dfrac{1}{V_{\textrm{NP}}}\int_{V_{\textrm{NP},i}} p_{\textrm{abs}}({\bf r}) d^3 r = \dfrac{1}{V_{\textrm{NP}}}\int \int_{V_{\textrm{NP},i}} \dfrac{\omega}{2}\varepsilon_m^{\prime\prime}|{\bf E}(\omega,{\bf r})|^2 d^3 r d\omega\nonumber\\ &= \dfrac{1}{V_{\textrm{NP}}} \int i_{\textrm{inc}}(\omega)e^{- z_i/\delta_{\textrm{skin}}(\omega)} \sigma_{\textrm{abs}}(\omega) d\omega,
\end{align}
where
$$\int_{V_{\textrm{NP},i}} \dfrac{\omega}{2}\varepsilon_m^{\prime\prime}|{\bf E}(\omega,{\bf r})|^2 d^3 r = i_{\textrm{inc}}(\omega)e^{-z_i/\delta_{\textrm{skin}}(\omega)} \sigma_{\textrm{abs}}(\omega).$$

Furthermore, since the heat equation~(\ref{eq:general_heat_eq_steady}) is a linear differential equation, the temperature $T({\bf r})$ in the multiple NP problem can be written as the linear combination of all the {\it single} NP temperature contributions (denoted by $\Delta T_i({\bf r})$) under CW illumination~\cite{thermo-plasmonics-multi_NP}, namely,
\begin{align}\label{eq:heat_eq_sng_NP_cw_sol}
\Delta T_i({\bf r}) = \dfrac{V_{\textrm{NP}}\bar{p}_{\textrm{abs,}i}}{4\pi\kappa_h} \begin{cases}
1/R, & \textrm{for } |{\bf r}-{\bf r}_i|<R,\\
1/|{\bf r}-{\bf r}_i| & \textrm{for } |{\bf r}-{\bf r}_i|>R,
\end{cases}
\end{align}
where $\bar{p}_{\textrm{abs,}i}$ is the absorbed power density by the {\it single} particle, given by Eq.~(\ref{eq:avg_pabs_i}); the symbol $\Delta$ denotes the difference with respect to the temperature in the absence of illumination, $T_{\textrm{dark}}$. Then, the solution for the multiple NP problem is
\begin{align}\label{eq:mti_NP_sum}
\Delta T({\bf r}) = \begin{cases}
\dfrac{V_{\textrm{NP}}}{4 \pi \kappa_h}\left[\dfrac{\bar{p}_{\textrm{abs},i}}{R}+\displaystyle\sum\limits_{j\neq i}\dfrac{\bar{p}_{\textrm{abs},j}}{|{\bf r}_j -{\bf r}_i|}\right],&\textrm{for NP at } {\bf r}_i,\\
\hfil\dfrac{V_{\textrm{NP}}}{4 \pi \kappa_h }\displaystyle\sum\limits_{j} \dfrac{\bar{p}_{\textrm{abs},j}}{|{\bf r}_j -{\bf r}_i|},& \textrm{ for } {\bf r} \textrm{ in the host,}
\end{cases}
\end{align}
The summation can be converted into an equivalent integration by dividing by the NP density~\cite{thermo-plasmonics-multi_NP}, namely, \begin{align}\label{eq:sum_2_int_cw}
\sum_{j\neq i}\dfrac{e^{-z_j/\delta_{\textrm{skin}}}}{|{\bf r}_j - {\bf r}_i|} \rightarrow \dfrac{1}{p^3} \int_{V^{\prime}_{\textrm{compsite}}} \dfrac{e^{-z^{\prime}/\delta_{\textrm{skin}}}}{|{\bf r}^{\prime} - {\bf r}_i|} d^3 r^{\prime},
\end{align}
where $V^{\prime}_{\textrm{compsite}}$ denotes the composite volume under the illumination but without the unit cell at ${\bf r}_i$.

Once the temperature is determined, one can define the photo-thermal conversion coefficient $a$, namely,
\begin{align}\label{eq:def_pht_thm_conver_coeff_Linic}
a = \dfrac{\langle\Delta T \rangle_{\textrm{top surface}}}{\int i_{\textrm{inc}}(\omega)d\omega},
\end{align}
where $\langle\Delta T \rangle_{\textrm{top surface}} = \int_{\textrm{top surface}}\Delta T({\bf r})\rho d\rho d\phi/A$ is the average temperature over the top surface of the pellet~\footnote{We average the temperature over the surface only because the reaction occurring within the skin depth of the pellet has the highest reaction rate due to the highest temperature and will mainly contribute to the overall increase of the reaction rate.}.

We first calculate the absorption cross-section of a single NP by using the Ag permittivity from the reference~\cite{Shalaev_ellipsometry_silver} and Mie theory; the result is shown in Fig.~\ref{fig:ext_abs_crx_sec_ag_nsph}. One can see that the scattering dominates absorption; a quadrupole resonance is seen at $\lambda \sim 380$ nm and a broad dipole resonance is seen at $\lambda \sim 480$ nm. The latter dominates the particle absorption since it overlaps with the spectrum of the CW light source much better than the quadrupole resonance.

\begin{figure}[h]
    \centering
    \includegraphics[width=8truecm]{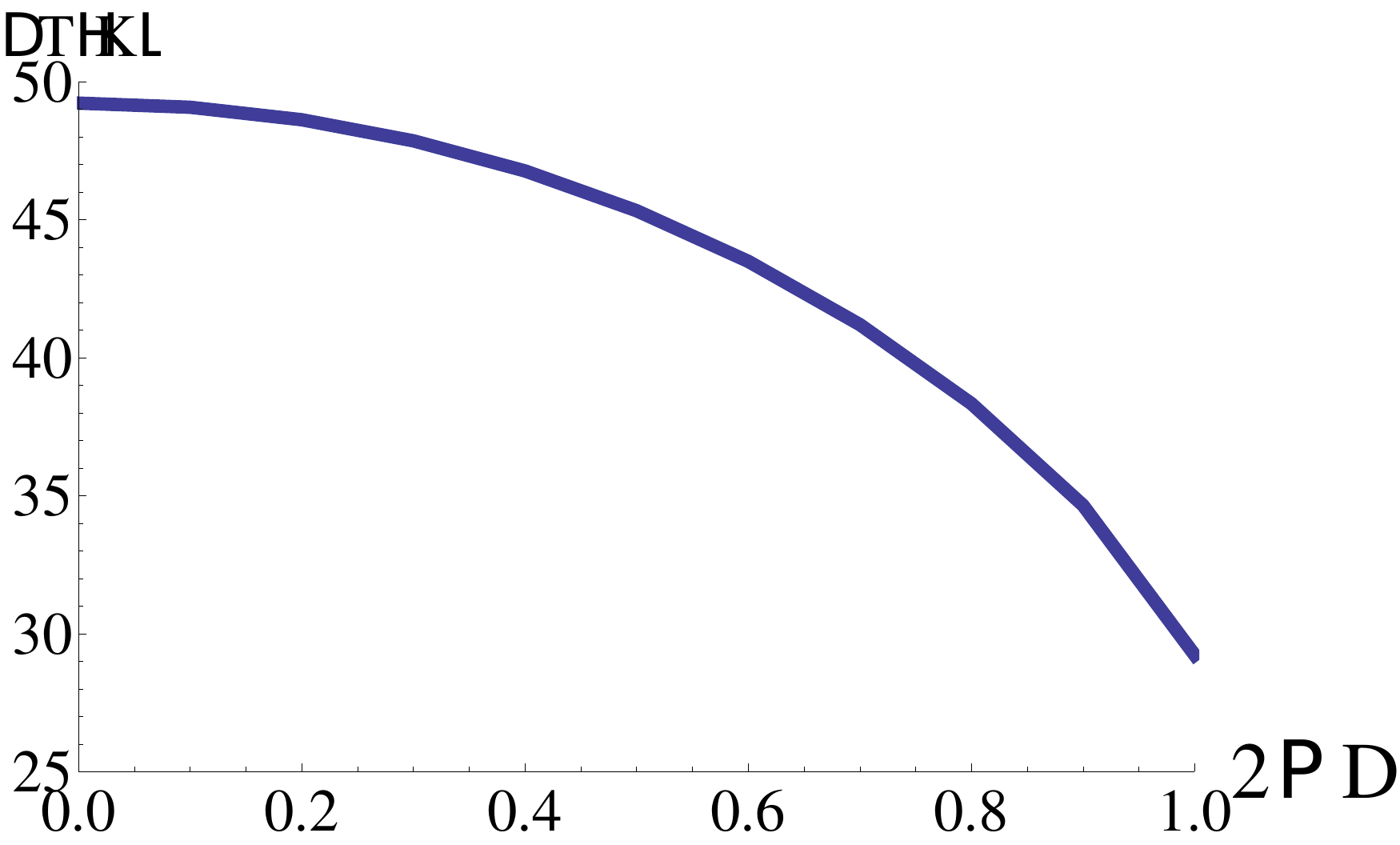}
    \llap{\raisebox{2cm}{\includegraphics[height=3truecm]{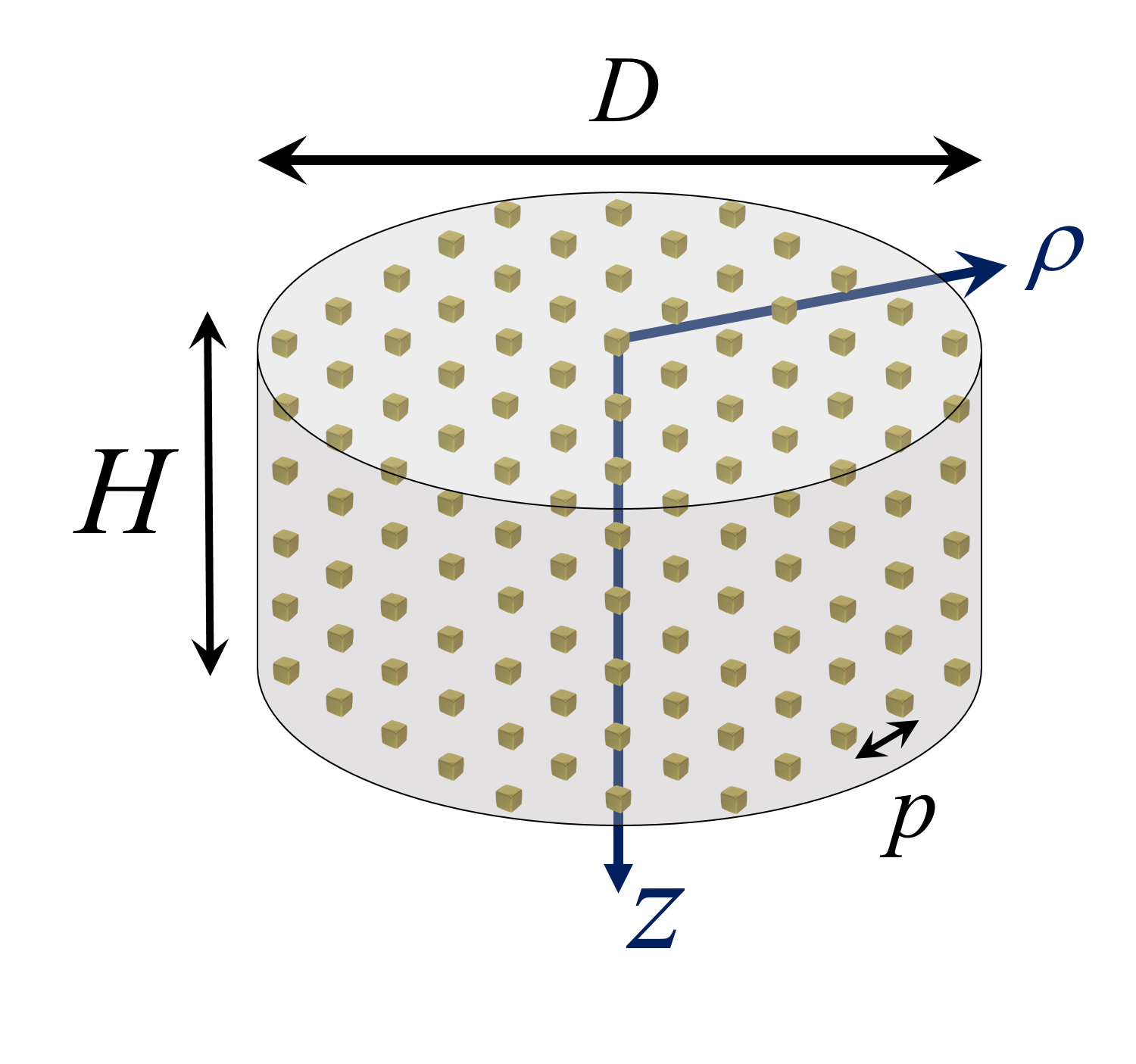}}}
    \caption{(Color online) Temperature rise of the top surface ($z = 0$) of the pellet under CW illumination of intensity 1 W/cm$^2$, $\rho$ is the radius coordinate and $D$ is the diameter of the pellet, as shown in the inset.}
    \label{fig:dtemp_Linic_top_surf_1Wcm-2}
\end{figure}

Then, the temperature profile on the top surface of the pellet can be obtained by Eq.~(\ref{eq:mti_NP_sum}), see Fig.~\ref{fig:dtemp_Linic_top_surf_1Wcm-2}. One can see that the temperature of the top surface decreases gradually from 50 K at the center ($\rho = 0$, $z = 0$) to 30 K at the edges ($\rho = D/2$, $z = 0$). Thus, the overall composite temperature rise is, in fact, a many-particle effect, much higher than the temperature rise (4.3 mK) of a single-particle model used in reference~\cite[paper III]{plasmonic_photocatalysis_Linic} and reference~\cite{christopher2011visible}. After averaging the surface temperature, Eq.~(\ref{eq:def_pht_thm_conver_coeff_Linic}) gives $a = 40$ K cm$^2$/W. This value is within the range of the value obtained for $\tilde{a}$ from the shifted Arrhenius Law in Sec.~\ref{sec:analysis_III} (Fig.~\ref{fig:Linic1}).

One should bear in mind that the value we obtained for $a$ should be considered only as an order-of-magnitude estimate. Indeed, as mentioned above, the authors of paper III did not indicate the beam size. For example, if the beam radius is changed from 5.6 mm to 2.8 mm, the maximum composite temperature rise will change from 50 K to 24 K for illumination intensity of 1 W/cm$^2$. The value for $a$ would also depend on the inter-particle separation. Specifically, larger inter-particle separation may lead to a decrease in the composite temperature. However, this will also cause a increase in the skin depth (see Eq.~(\ref{eq:skin_depth})), hence to broadening of the heat source. Yet, since the skin depth in~\cite[paper III]{plasmonic_photocatalysis_Linic} is much smaller than the thickness of the pellet, still all the incident photon energy is absorbed. Numerical solutions of Eq.~(\ref{eq:general_heat_eq_steady}) show that the pellet temperature rise is only weakly sensitive ($< 5\%$) to  inter-particle separation (via the the thickness of the heat source) in the range of 200 nm $<p<$ 400 nm (which is the inter-particle range deduced from SEM pictures in paper III).



In addition, one might expect that the approximation of the nanocube by a nanosphere will yield a somewhat different value for $a$ due to the difference in the respective absorption cross-sections. However, similarly, an increase of $\sigma_{\textrm{abs}}$ also causes a decrease of skin depth (see Eq.~(\ref{eq:skin_depth})) such that the composite temperature rise is also weakly sensitive to the particle shape.


\subsection{Detailed temperature calculations for paper IV}\label{app:Halas_IV}
The catalyst pellet in~\cite[paper IV]{Halas_Science_2018} consisted of Cu-Ru NPs of radius $2.5$ nm supported on larger porous Al$_2$O$_3$-MgO particles. We approximate the system as a Cu-Ru nanosphere (periodic) array immersed in a uniform host material. Under these assumptions, one can deduce from the measurements of the Cu concentration reported in reference~\cite[paper IV]{Halas_Science_2018} that the average inter-particle separation is $p = 24.5$ nm. Further, the optical properties of the metal NPs are characterized by $\varepsilon_m = 0.99 \varepsilon_{\textrm{Cu}} + 0.01\varepsilon_{\textrm{Ru}}$~\footnote{In that respect, the illustration in IV is somewhat misleading - Ru is far more scarce than Cu.} according to the element concentration measurements in~\cite[paper IV]{Halas_Science_2018}; since the thermal properties are similar for Cu and Ru, they are assumed simply to be $c_m = c_{\textrm{Cu}}$, $\rho_m = \rho_{\textrm{Cu}}$ and $\kappa_m = \kappa_{\textrm{Cu}}$.

For the host material, we set the host permittivity to be $\epsilon_h = (1 - f_{\textrm{v}})\varepsilon_{\textrm{air}} + 0.5 f_{\textrm{v}}\varepsilon_{\textrm{Al}_2\textrm{O}_3} + 0.5f_{\textrm{v}}\varepsilon_{\textrm{MgO}}$, the host volumetric heat capacity to be $\rho_h c_h = (1 - f_{\textrm{v}}) \rho_{\textrm{air}} c_{\textrm{air}} + 0.5 f_{\textrm{v}} \rho_{\textrm{Al}_2\textrm{O}_3} c_{\textrm{Al}_2\textrm{O}_3} + 0.5 f_{\textrm{v}} \rho_{\textrm{MgO}} c_{\textrm{MgO}}$ and the thermal conductivity of the host to be~\cite{birdtransport,Pietrak-eff-kappa-compos-2014} 
\begin{align}\label{eq:eff_kapa_h_halas}
\kappa_h = \kappa_{\textrm{air}}+\dfrac{3f_{\textrm{v}}\kappa_{\textrm{air}}}{\frac{(\kappa_{\textrm{Al}_2\textrm{O}_3} + \kappa_{\textrm{MgO}})/2 + 2\kappa_{\textrm{air}}}{(\kappa_{\textrm{Al}_2\textrm{O}_3} + \kappa_{\textrm{MgO}})/2 - \kappa_{\textrm{air}}} - f_v},
\end{align}
where $f_{\textrm{v}}$ is the volume fraction of oxides in the composite which can be deduced from the mass (1.1 mg) and the volume (diameter $D = $ 2 mm and thickness $H = $ 1 mm) of the pellet~\cite{Halas_Science_2018}. As in App.~\ref{app:Linic}, the relatively more advanced effective medium formula used for the thermal conductivity is required due to the large differences between the conductivities of the constituents. This choice for the various filling factors is confirmed by comparing the calculated absorption cross-section $\sigma_{\textrm{abs}}$ shown in Fig.~\ref{fig:cu_ru_abs_crx_sec} with the diffusive reflection measurement shown in the supplementary information of reference~\cite[paper IV]{Halas_Science_2018}.

\begin{figure}[h]
\centering
\includegraphics[width=6truecm]{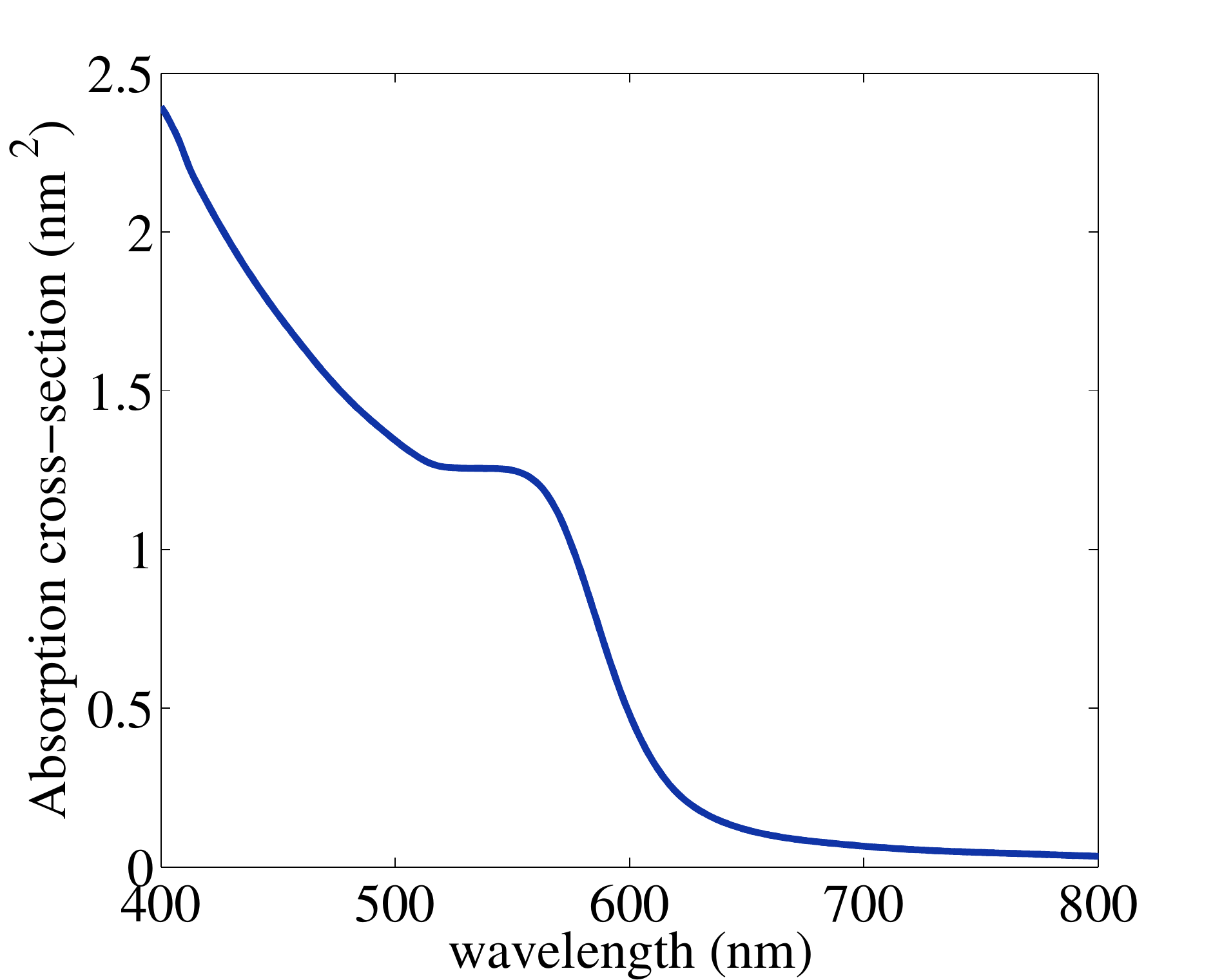}
\caption{(Color online) Calculated absorption cross-section of Cu-Ru NP ($\varepsilon_m = 0.99\epsilon_{\textrm{Cu}} + 0.01\epsilon_{\textrm{Ru}}$) in a uniform host (($\varepsilon_h = 0.9\epsilon_{\textrm{air}} + 0.05\epsilon_{\textrm{Al}_2\textrm{O}_3} + 0.05\epsilon_{\textrm{MgO}}$).}\label{fig:cu_ru_abs_crx_sec}
\end{figure}

The pulse train illumination on the pellet is set below to have a (time) average intensity $\langle I \rangle = $ 3.2 W/cm$^2$ with central wavelength 550 nm (as in Fig.~\ref{fig:IV_1}(A) in Sec.~\ref{sec:analysis_IV}), spot size $A = \pi \cdot 1$ mm$^2 \sim D^2$, average power $\langle P \rangle = \langle I \rangle A$, pulse repetition rate $f = $ 80 MHz (i.e., pulse period $\sim 12.5$ ns) and pulse duration $\tau = 4$ ps. The energy per pulse is thus $\langle P \rangle/f$, and the peak pulse intensity is $I_{\textrm{max}} = \langle P \rangle/(A \cdot f \cdot \tau) = \langle I \rangle/(f \cdot \tau)$.

\subsubsection{Temperature dynamics of the composite - formulation}
The spatio-temporal evolution of the catalyst temperature, $T({\bf r},t)$, can be determined by solving the heat equation,
\begin{align}\label{eq:general_heat_diffuse_NP_array}
\begin{cases}
\rho_m c_m \dfrac{\partial T({\bf r},t)}{\partial t} - \kappa_m \nabla^2 T({\bf r},t) = p_{\textrm{abs}}({\bf r},t), &\textrm{for } {\bf r} \textrm{ in NPs,} \\
\hfil \rho_h c_h \dfrac{\partial T({\bf r},t)}{\partial t} - \kappa_h \nabla^2 T({\bf r},t) = 0,  &\textrm{for } {\bf r} \textrm{ in the host,}
\end{cases}
\end{align}
with appropriate boundary conditions at the surface of each NP~\cite{Baffou_pulsed_heat_eq_with_Kapitza}. 
For simplicity (as in App.~\ref{app:Linic}), we ignore the temperature dependence of $\kappa_h$, $c_h$ and $\rho_h$. This dependence should be included for sufficiently high temperatures, typically, for $T > 400$ K forcing one to solve Maxwell's equations together with the heat equation~(\ref{eq:general_heat_eq_steady}) self-consistently~\cite{Sivan-Chu-high-T-nl-plasmonics,Gurwich-Sivan-CW-nlty-metal_NP}. This is motivated by the weak sensitivity of the pellet temperature distribution to the inter-particle separation and absorptivity, as discussed in App.~\ref{app:Linic}.

As in App.~\ref{app:Linic}, due to the absorption by the NPs, the peak pulse intensity drops along the thickness of the pellet. Since $R \ll p \ll \lambda$ and absorption dominates scattering for NP size of a few nm, we can apply the effective medium approximation such that the spatial dependence of the peak pulse intensity can be written as $ I_{\textrm{max}} \exp(-z/\delta_{\textrm{skin}})$. The penetration (skin) depth is estimated by
\begin{align}
\delta_{\textrm{skin}} = p^3/\sigma_{\textrm{abs}} \sim 12 \ \mu \textrm{m},
\end{align}
a value which is similar to that (10 $\mu$m) provided in the supplementary of reference~\cite[paper IV]{Halas_Science_2018}.

Since the heat equation is a linear differential equation, the problem can be simplified by first looking for the temperature evolution of a {\it single} NP at ${\bf r} = 0$ under a {\it single} pulse illumination at $t = 0$, denoted by $\Delta T_{0,0}({\bf r,t})$, namely,
\begin{align}\label{eq:general_heat_diffuse_sng_NP_sng_pulse}
\begin{cases}
\rho_m c_m \dfrac{\partial T_{0,0}({\bf r},t)}{\partial t} - \kappa_m \nabla^2 T_{0,0}({\bf r},t) = p_{\textrm{abs},0,0}({\bf r},t), & \textrm{for } r < R, \\
\hfil \rho_h c_h \dfrac{\partial T_{0,0}({\bf r},t)}{\partial t} - \kappa_h \nabla^2 T_{0,0}({\bf r},t) = 0, & \textrm{for } r > R,
\end{cases}
\end{align}
with appropriate boundary conditions at $r = R$~\cite{Baffou_pulsed_heat_eq_with_Kapitza}, and where $p_{\textrm{abs},0,0}({\bf r},t)$ is the absorbed power density under a {\it single} pulse illumination, the integration of which over space-time is the total energy absorbed per pulse by a {\it single} NP, $\int\int p_{\textrm{abs},0,0}({\bf r},t) d{\bf r} dt = \mathcal{E}_0 = \sigma_{\textrm{abs}}\langle I \rangle/f$. Then, the solution for the pulse train illumination of the multiple particle composite can be obtained by the linear combination of many solutions of {\it single} pulse events from all particles, namely,
\begin{align}\label{eq:cps_temp_sum}
\Delta T({\bf r},t) = \sum\limits_{t_k<t}\sum\limits_{j} \Delta T_{0,0}({\bf r} - {\bf r}_j,t - t_k)\exp(-z_j/\delta_{\textrm{skin}}),
\end{align}
where $t_k = k/f$ is the pulse time, $k = 0,1,\dots$. Eventually, the system reaches a ``steady-state'' (see Fig.~\ref{fig:dtemp_sng_pulse_mti_pulse_dyn}(B)), as shown below, in which case the photo-thermal conversion coefficient can be defined by
\begin{align}\label{eq:def_pht_thm_conver_coeff_halas}
a = \dfrac{\langle \Delta T(t\rightarrow\infty) \rangle_{\textrm{top surface}}}{\langle I \rangle},
\end{align}
where $\langle \Delta T(t\rightarrow\infty) \rangle_{\textrm{top surface}}$ stands for the average temperature on the top surface of the pellet in the ``steady-state''.

In what follows, we discuss these calculation steps separately.

\subsubsection{Single particle temperature $\Delta T_{0,0}$}\label{app:sng_NP_temp}
The spatio-temporal evolution of the NP temperature is a result of a series of processes: {\it 1}. the inner temperature rise dynamics within the NP (due to photon absorption) occurring on a time scale of the pulse duration $\tau$, and resulting in an increase of inner temperature by $\mathcal{E}_0 / \rho_m c_m V_{\textrm{NP}} \approx 1.8$ mK, {\it 2}. the inner NP temperature decay due to heat transfer to the host, estimated to occur within $\tau^{\textrm{d}}_{\textrm{NP}} \equiv R^2 \rho_m c_m / 3 \kappa_h \sim 0.2$ ns~\cite{Baffou_pulsed_heat_eq_with_Kapitza} and {\it 3}. heat diffusion in the host, occurring on a much longer time scale. 
Since we are interested only in the long time dynamics (specifically, the ``steady-state'' of the composite temperature) for the purpose of the photocatalysis experiments and since $\tau \ll \tau^{\textrm{d}}_{\text{NP}} \ll 1/f$, we can treat these stages separately without compromising the accuracy. Furthermore, the heat equation~(\ref{eq:general_heat_diffuse_sng_NP_sng_pulse}) can be simplified by approximating the NP as a point-source such that the absorbed power density is represented by a space-time Dirac delta distribution~\cite{Baffou_pulsed_heat_eq_with_Kapitza},
\begin{align}\label{eq:heat_diffus_sng_NP_sng_pulse_delta}
\rho_h c_h \dfrac{\partial T_{0,0}({\bf r},t)}{\partial t} = \kappa_h \nabla^2 T_{0,0}({\bf r},t) + \mathcal{E}_0\delta({\bf r})\delta(t).
\end{align}
Eq.~(\ref{eq:heat_diffus_sng_NP_sng_pulse_delta}) has the analytic solution
\begin{align}\label{eq:heat_diffus_delta_sng_NP_sng_pulse_sol}
\Delta T_{0,0}({\bf r},t) = \dfrac{\mathcal{E}_0}{\rho_h c_h}\dfrac{1}{(4\pi d_h t)^{3/2}} \exp\left(-\dfrac{r^2}{4 \pi d_h t}\right),
\end{align}
where $d_h = \kappa_h/(\rho_h c_h)$ is the diffusivity of the host.

\subsubsection{The steady-state temperature and temperature uniformity under pulse train illumination}
\begin{figure}[h]
\centering
\includegraphics[width=7truecm]{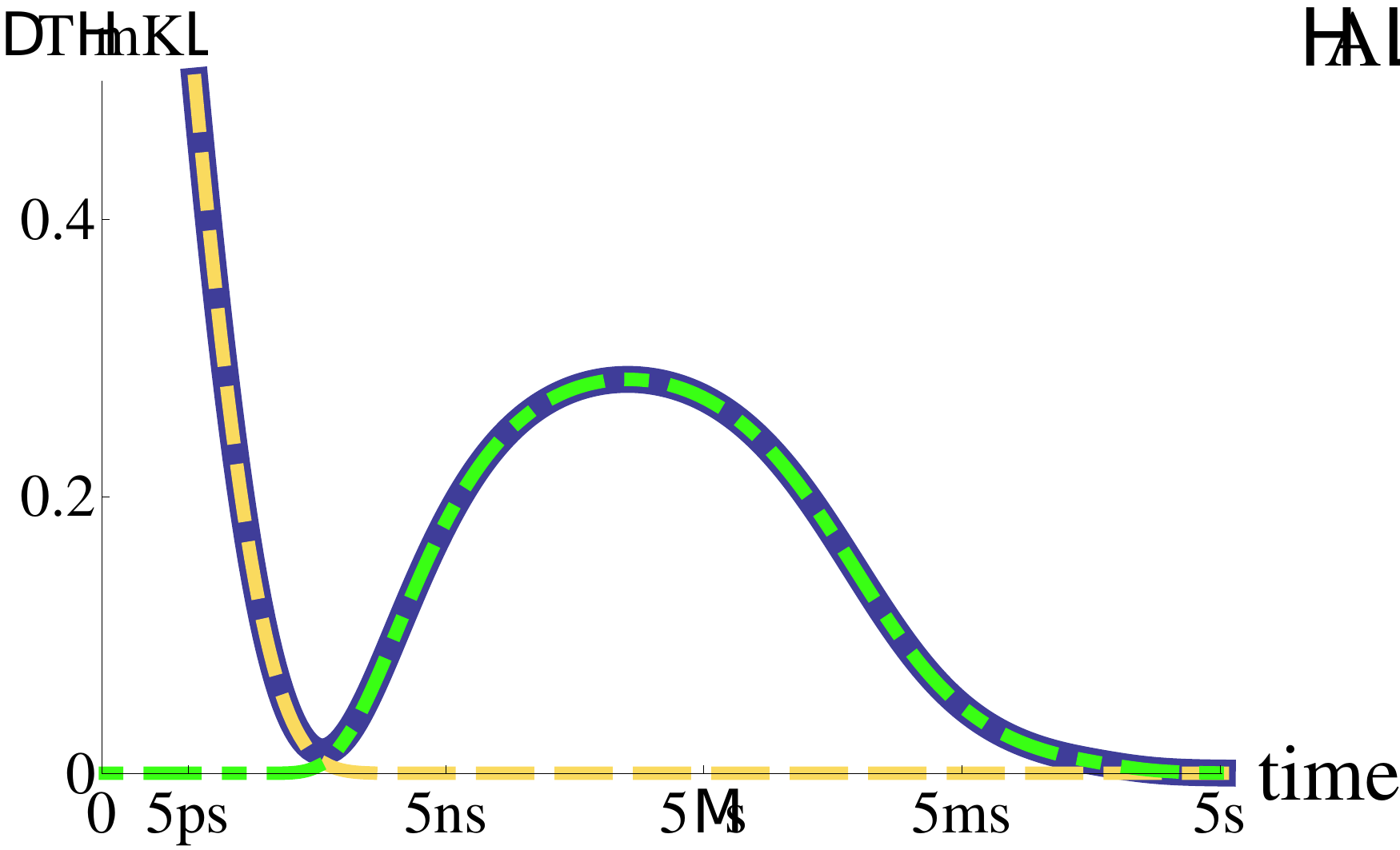}
\includegraphics[width=7truecm]{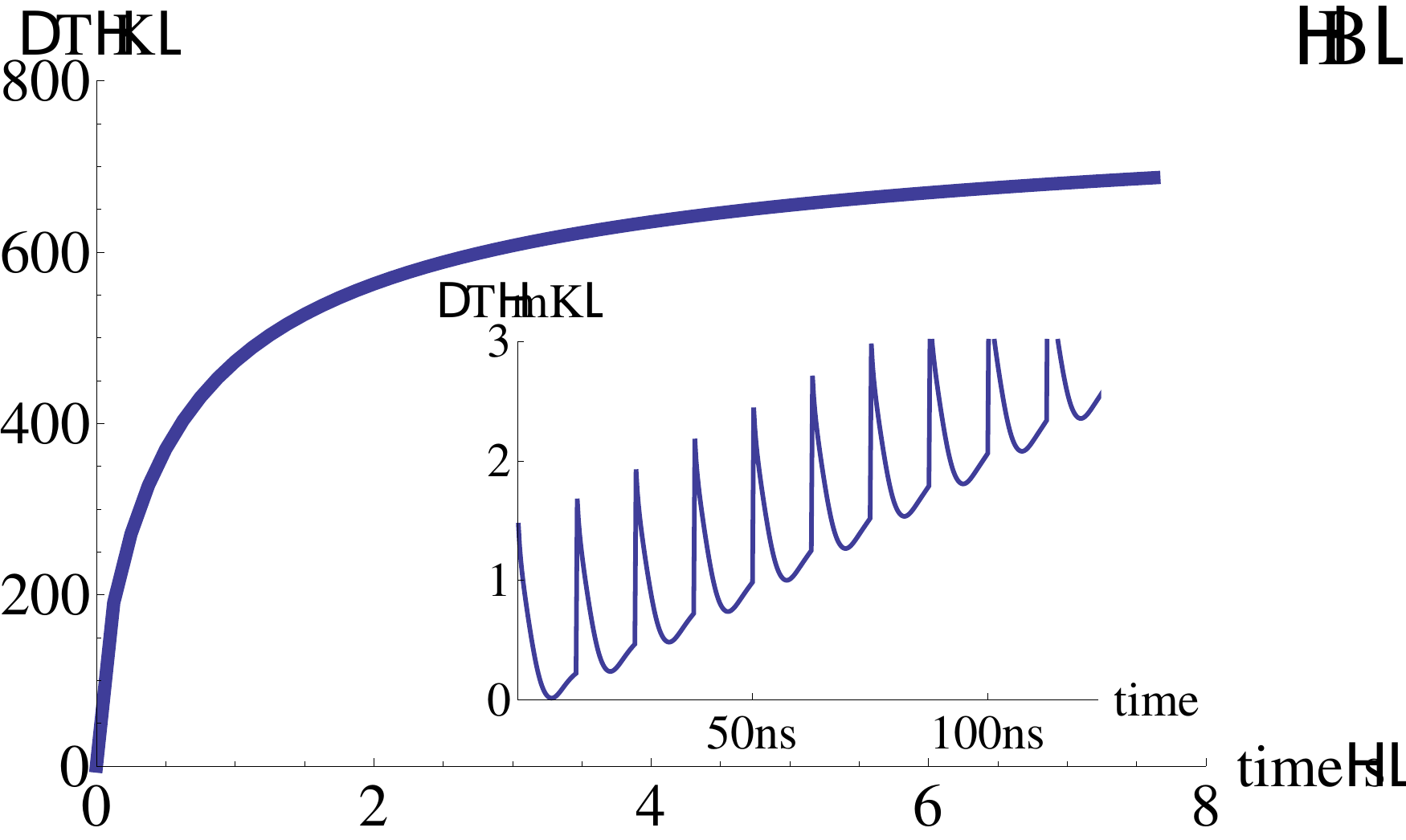}
\caption{(Color online) (A) Temperature evolution at the top center of the pellet under a {\it single} pulse illumination (blue solid line), the contribution from the inner temperature rise (yellow dashed line) and the heat diffusion from (many) other NPs (green dash-dotted line) are also shown. Note that the time axis is not a linear scale. (B) Temperature evolution at the top center of the pellet under {\it  pulse train} illumination. The insert shows the temperature evolution during the illumination of the first several pulses, the time axis is similar to that in (B).}
\label{fig:dtemp_sng_pulse_mti_pulse_dyn}
\end{figure}

In order to understand the temperature evolution under {\it pulse train} illumination, we first study the temperature evolution of the NP at the top center of the pellet for a {\it single} pulse illumination, see Fig.~\ref{fig:dtemp_sng_pulse_mti_pulse_dyn}(A). One can see that most of the absorbed energy leaves the NP and diffuses in the host so that the inner temperature decays within the first 2 ns. Then, the NP temperature increases due to the heat diffusion from the (many) other NPs, such that this heat diffusion keeps the NP warm at 0.25 mK for more than 50 ms. Eventually, the NP temperature decays again to zero when all the thermal energy diffuses out of the pellet.

\begin{figure}[h]
\centering
\includegraphics[width=7truecm]{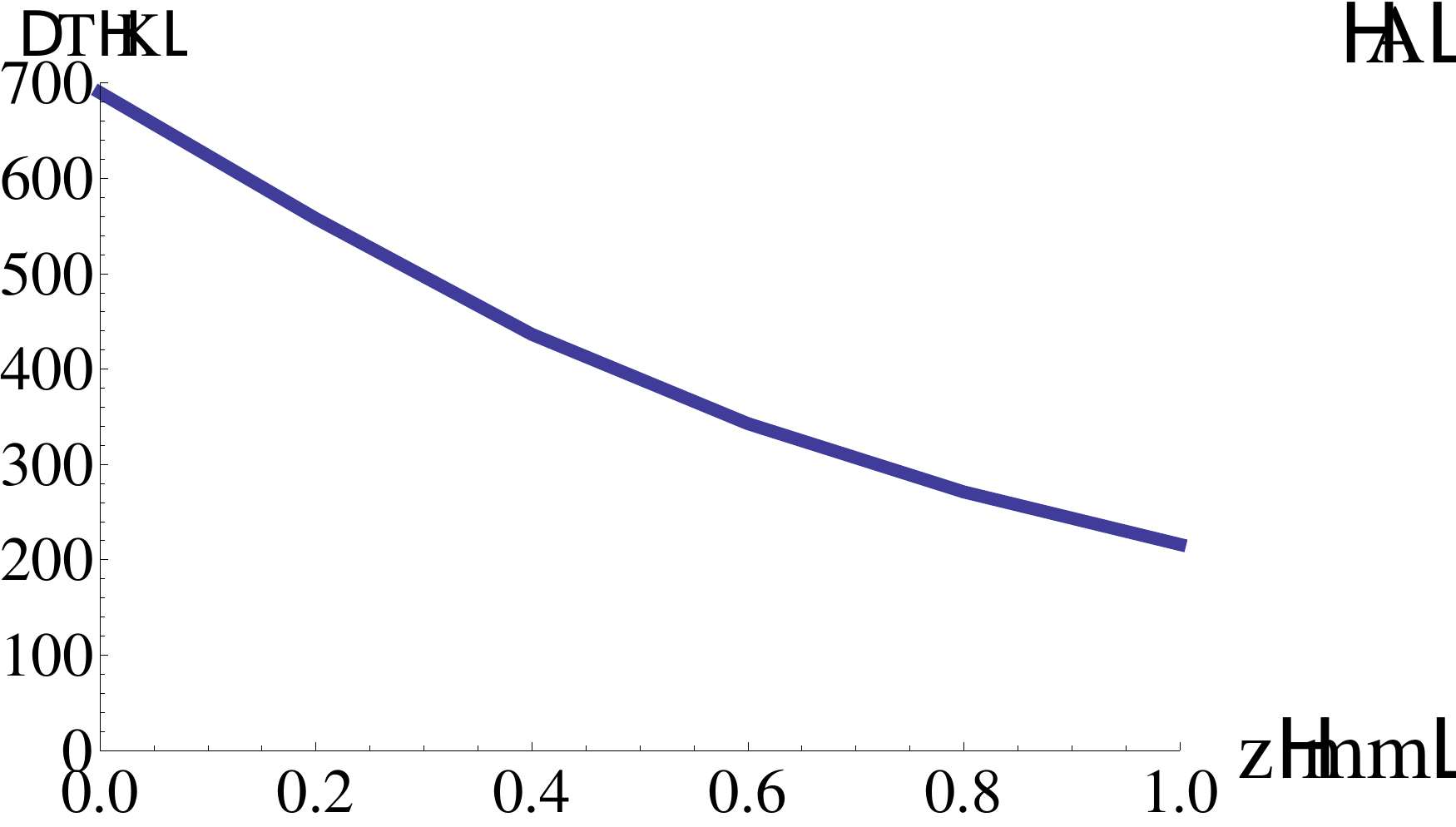} \includegraphics[width=7truecm]{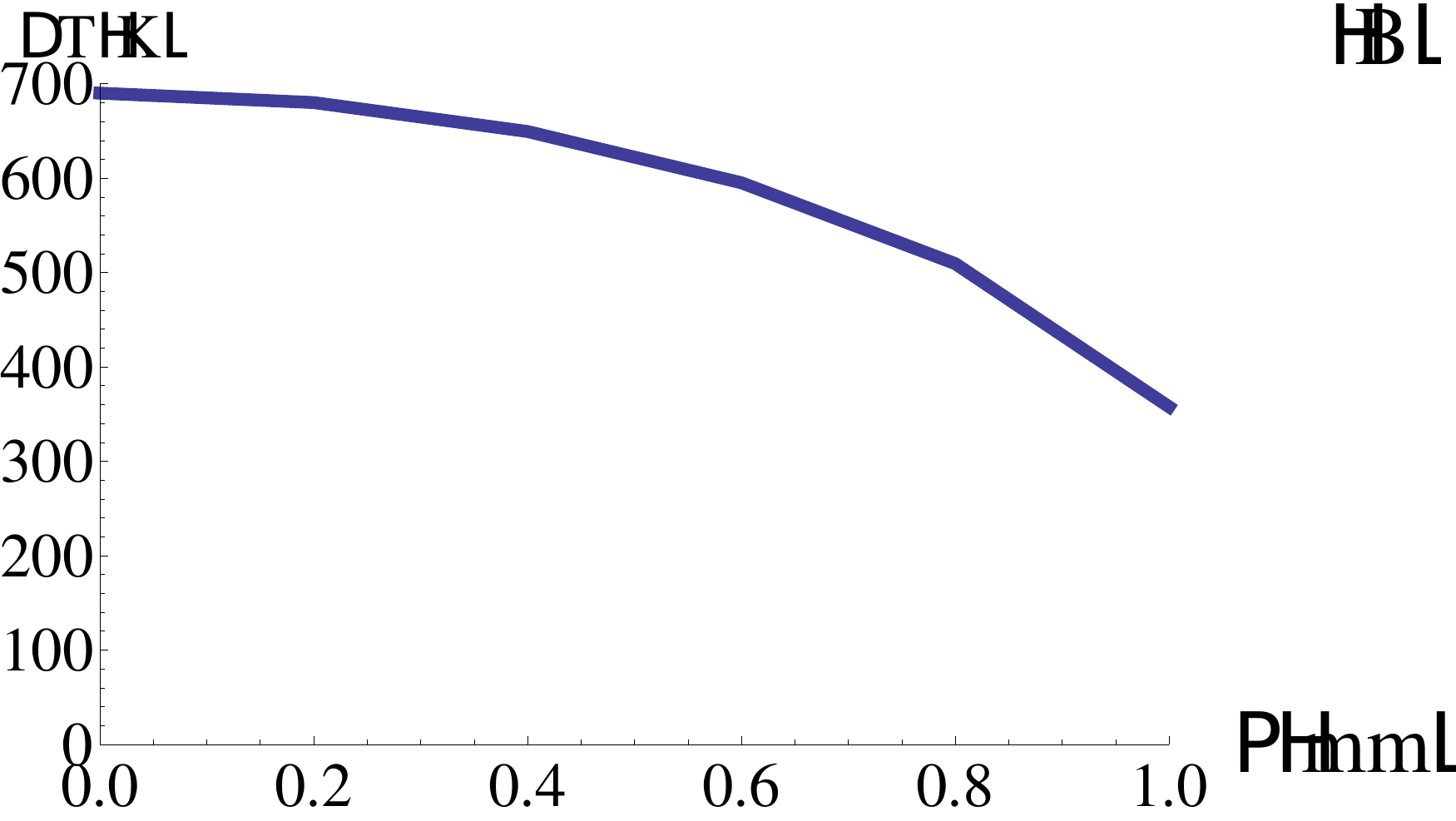}
\caption{(Color online) Temperature rise profile (A) along the cylindrical axis of the pullet and (B) along the radial direction on the top surface of the pellet.} \label{fig:temp_uniformity_mti_pulse}
\end{figure}

Then, the temperature evolution under {\it pulse train} illumination can be obtained by a summation of many (time-shifted) {\it single} pulse events. Since the pulse repetition rate is faster than the overall decay time to the environment, there is an overall (``step-wise'') temperature buildup under {\it pulse train} illumination, see Fig.~\ref{fig:dtemp_sng_pulse_mti_pulse_dyn}(B). This heat accumulation finally slows down and the temperature reaches a ``steady-state'' of $\Delta T \sim 680$ K on a time scale of a few seconds, as shown in Fig.~\ref{fig:dtemp_sng_pulse_mti_pulse_dyn}(B). One can also calculate the ``steady-state'' temperature profile on the top surface of the pellet, see Fig.~\ref{fig:temp_uniformity_mti_pulse}(B). The photo-thermal conversion coefficient is deduced to be $\sim$ 170 K cm$^2$/W in Sec.~\ref{sec:analysis_IV}.

Similar to the analysis in App.~\ref{app:Linic}, the ``steady-state'' temperature is weakly sensitive to the inter-particle separation (within the parameter range deduced from the SEM pictures) 
due to the opposite effect of the particle density and the skin depth on the ``steady-state'' temperature. We should note that our calculations assumed for simplicity that the chamber in which the pellet is held has infinite size. In practice, the actual pellet temperature could be partially reduced by a few ten percents because the temperature of the chamber walls was maintained at 300 K in the experiment~\cite{Halas_Science_2018}. Due to all the above, overall, the value obtained for the photo-thermal conversion coefficient should be viewed as an order-of-magnitude estimate. 

Having said the above, we should emphasize that the most important aspect of our calculation is qualitative, as it shows the significant temperature gradients across the pellet. Indeed, since only the NPs in the pellet surface layer of thickness ($\sim \delta_{\textrm{skin}}$) generate heat under illumination, large temperature non-uniformity would be expected across the pellet. By using Eqs.~(\ref{eq:cps_temp_sum}) and~(\ref{eq:heat_diffus_delta_sng_NP_sng_pulse_sol}), we can calculate the ``steady-state'' temperature profile along the cylindrical axis and along the radial direction on the top surface of the pellet, see Figs.~\ref{fig:temp_uniformity_mti_pulse}(A) and (B), respectively. One can see that the temperature gradually decreases from 680 K to 250 K along the cylindrical axis. In the transverse direction, the non-uniformity is somewhat smaller. As explained in Sec.~\ref{sec:discussion} (and~\cite{anti-Halas_comment}), these non-uniformities cause severe differences between the thermal contributions in the photocatalysis and thermocatalysis control experiments, thus, invalidating the conclusions of~\cite{Halas_Science_2018}.

\section{A thermal interpretation of the kinetic isotope effect of paper III}\label{app:KIE_analy}
In paper III, the authors present results of a kinetic isotope effect (KIE), which they interpret as a hallmark of the ``hot''-electron mechanism. Here, we show that an excellent fit to the KIE can be made using only a thermal effect, i.e., the shifted-Arrhenius theory. To understand this, we briefly reiterate the KIE experiments as described in III. The experiment is done by measuring the reaction rate (O$_2$ dissociation in ethylene epoxidation) using both the abundant isotope ($^{16}$O) and the rare isotope ($^{18}$O), as a function of illumination intensity. At each illumination intensity, the external (i.e., measured) temperature is reduced (from an initial value of $498$K) such that the reaction rate of the $^{16}$O dissociation remains constant (interestingly, such a measurement could have been used to estimate the relation between measured and effective temperatures). The experiments start at 498K and temperatures are reduced from there.

Within our theoretical description, the effective temperature of the catalyst (i.e., the temperature felt by the reaction) is then $T = 498 + a I_{inc} - \Delta T$
, where $\Delta T$ is the change in the measured temperature from the initial measurement (performed in the dark, i.e., $I_{inc} = 0$. The reaction rate of $^{16}$O$_2$ dissociation is then
\begin{equation}
R_{16} = R_0 \exp \left( -\frac{E_a}{k_B (498 + a I_{inc} - \Delta T} \right) = \mathrm{constant}.
\end{equation}

The reaction rate of $^{18}$O$_2$ obeys a similar formula, with two differences in parameters. The first is the (slightly) different activation energy, which is the cause of the (dark) isotope effect, i.e., the difference in the reaction rates at $I_{inc} = 0$. For Oxygen, the KIE was $1.09$, such that the difference in the activation energies can easily be computed (not fitted), and is found to be $0.00369$eV (this is a small difference, in fact smaller than the measured temperature).

The second difference is that the measurement can have a slightly different photo-thermal conversion coefficient $a_{18}$. As demonstrated above, the photo-thermal conversion coefficient is very sensitive to the details of the sample. If $a_{18} \neq a_{16}$, then, when the reaction temperature is reduced to keep $R_{16}$ constant, the temperature of the isotope reaction is in fact changing with illumination.

\begin{figure}[h]
\centering
\includegraphics[width=7truecm]{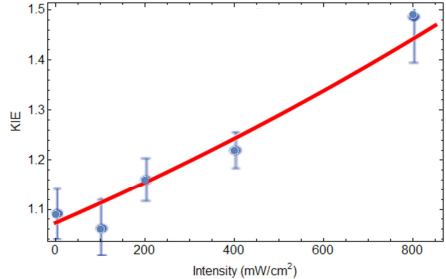}     \caption{The kinetic isotope effect data of paper III (blue points) and the fit to an Arrhenius theory (solid red line). Obtaining this fit requires a single fitting parameter, which is the photo-thermal conversion factor for the rare isotope measurement. The data can be fitted to excellent degree with a very small difference in the photo-thermal conversion coefficient, $a_{18}/a_{16} = 0.016$, i.e., around 1.5\%. } \label{fig:KIE}
\end{figure}

In Fig.~\ref{fig:KIE}, we plot the KIE data taken from paper III (blue points, including the error bars). On top of that, we plot the theoretical KIE (solid red line), which is simply
\begin{eqnarray}
KIE &=& R_{16}/R_{18}~~,  \nonumber \\
R_{18} &=& R_0 \exp \left( -\frac{E_a}{k_B (498 + a_{18} I_{inc} - \Delta T)} \right) \nonumber \\
&=& R_0 \exp \left( -\frac{E_a}{k_B (498 + (a_{18} - a_{16}) I_{inc})} \right)~.
\end{eqnarray}

To obtain the fit of Fig.~\ref{fig:KIE}, it turns out that one only needs a tiny change ($\sim 1.6 \%$) in the photothermal conversion coefficient, to reproduce the the experimental data. Thus, if $a_{16} = 0.36$  K cm$^2$/ W (taken from the fit to the data presented in the main text), then, the KIE can be fitted very well with $a_{18} = 0.366$ K cm$^2$/ W, as seen in Fig.~\ref{fig:KIE}.

\section{The effect of convection on the pellet temperature}\label{app:convection}
In the calculations of the previous sections, we assumed that the temperature distribution in the environment is determined just by heat diffusion. However, in references~\cite{Halas_dissociation_H2_TiO2,Halas_H2_dissociation_SiO2,plasmonic_photocatalysis_Linic,Halas_Science_2018}, it was attempted to remove some of the heat generated in the NPs by gas flows in the reaction chamber via convection. The movement and temperature of the gas should be rigorously described by the continuity equations for mass, momentum (Navier-Stokes equations) and energy~\cite{burmeister1993convective}. 
However, due to the large domain size, unknown parameters and geometry, such numerical calculation could be time-consuming and computational costly or even unfeasible.

Instead, let us estimate the heat power transferred via convection by assuming that it satisfies Newton's Law of cooling~\cite{burmeister1993convective}, i.e.,
\begin{align}\label{eq:newton_cool_law}
q_c = h_c \cdot A \cdot \Delta T,
\end{align}
where $q_c$ is the heat power transferred via convection, $h_c$ is the convective heat transfer coefficient, $A$ is the heat transfer surface area and $\Delta T$ is the difference between the pellet surface temperature and the temperature of the gas far away from the pellet.

The convective heat transfer coefficient is usually determined empirically because it depends not only on the properties of the catalyst and the gas, but also on the flow conditions and the inner geometry of the reactor. In the following estimate, for simplicity, we set $h_c =$ 22 W/(m$^2\cdot$K)~\cite{convec_Heat_transf}, the value as that of air with flow velocity of $2$ m/s~\cite{convec_Heat_transf}; this value is much higher~\footnote{A gas velocity of 2 m/s roughly corresponds to a gas flow rate of $\sim$ 4000 sccm - 12000 sccm for chamber cross-section area of $\sim$ 1 cm$^2$.} than those reported in references~\cite{Halas_dissociation_H2_TiO2,Halas_H2_dissociation_SiO2,plasmonic_photocatalysis_Linic,Halas_Science_2018}. 

In particular, in reference~\cite[paper I]{Halas_dissociation_H2_TiO2}, 10 sccm of H$_2$ and 10 sccm of D$_2$ were flown through the chamber, the area of the top surface of the pellet was $A \approx 16\pi$ mm$^2$ and the temperature difference is deduced to be $\Delta T \approx 65$ K for incident laser intensity of 2.4 W/cm$^2$ in Sec.~\ref{sec:analysis_I_II}. Thus, the heat power transferred via convection is $\sim$ 0.07 W, about $6$\% of the incident power. Similarly, in reference~\cite[paper II]{Halas_H2_dissociation_SiO2}, the photo-thermal conversion coefficient deduced in Sec.~\ref{sec:analysis_I_II} is $\sim 5 - 10$ times larger than that of reference~\cite[paper I]{Halas_dissociation_H2_TiO2}. Accordingly, the heat power transferred via convection is $\sim 5 - 10$ times larger, which is $\sim$ 30\% - 60\% of the incident power. In reference~\cite[paper III]{plasmonic_photocatalysis_Linic}, 20 sccm of ethylene along with 20 sccm O$_2$ and 60 sccm N$_2$ were flown through the chamber, the area of the top surface of the pellet was $A \approx$ 1 cm$^2$. The temperature difference is deduced to be $\Delta T \approx 32$ K for incident laser intensity of 800 mW/cm$^2$ in Sec.~\ref{sec:analysis_III} and App.~\ref{app:Linic}, so that the heat power transferred via convection is $\sim 26.4$ mW, about $9$\% of the incident power. Finally, in reference~\cite[paper IV]{Halas_Science_2018}, 5 sccm (100 sccm) of NH$_3$ for the intensity range of 1.6 W/cm$^2$ $-$ 3.2 W/cm$^2$ (4 W/cm$^2$ $-$ 9.6 W/cm$^2$) was flown through the chamber, the area of the top surface of the pellet was $A \approx \pi \cdot 1$ mm$^2$. In this case, for an illumination intensity of 3.2 W/cm$^2$, the heat power transferred via convection is $\sim$ 40 mW, which is $\sim$ 40\% of the incident power.

One can see that, at most, a few ten percents of the generated heat were removed by convection even for rates much higher than used in practice. Furthermore, since only the NPs within the skin depth act as heat sources and the skin depth is much shorter than the sample thickness, the temperature non-uniformity caused by the local heating is not expected to be eliminated by the convection.

\section{A discussion of the potential role of NP melting}\label{app:melting?}

In papers I - III, the temperature rise is moderate. However, in IV, higher illumination intensities were employed, such that the temperature rise was much higher; the linear relation (2)-(3) between the reaction temperature and the incident intensity even predicts temperatures in excess of 2000K. However, we should recall that at such high temperatures, the thermo-optic nonlinearity of the metal (and potentially of the host) causes the temperature to be much lower than the linear prediction~\cite{Sivan-Chu-high-T-nl-plasmonics,Gurwich-Sivan-CW-nlty-metal_NP}.
As to the question of melting, for small nanoparticles this is not a trivial issue. Indeed, it may not be precise to refer to the nanoparticles as being solid even at temperatures modestly above room temperature (e.g., for gold); instead they are unstable, in the sense that the atoms continuously migrate and the nanoparticle internal morphology fluctuates with time between various nearly degenerate states; these effects might depend, among other aspects, of the environment and to the best of our knowledge have not been characterized for Cu-Ru nanoparticles.

Nevertheless, let us adopt the severe assumption that the nanoparticles do undergo a well-defined phase transition, and furthermore that the melting temperature is significantly lower than the bulk melting temperature of Cu (but not of Ru, which has a far higher melting temperature). First and foremost, we must bear in mind that the temperature we extract from the experimental data is the maximum within the reactor. However, the calculations in App.~\ref{app:T_calculation} (as well as measurements reported in~\cite{Liu_thermal_vs_nonthermal,Meunier_T_uniformity,Liu-Everitt-Nano-Letters-2019}) show a rather significant temperature inhomogeneity inside the reactor such that the temperature of other parts of the reactor (most of its volume, in practice) are significantly lower, potentially by more than 50\%.

Thus, the question we should ask ourselves is how melting of the nanoparticles in a small part of the reactor affect the observed reaction rate. Overall, we tend to say that the effect will be, at most, rather small. This conjecture relies on several arguments:

1.  The antenna-reactor nanoparticles in the pellet were unstructured and perhaps fluxional even close to room temperature. Even if the melting causes shape modifications to each particle, this will have a small effect on average, and not affect the catalytic properties of the metal in any deleterious way.

2.  Due to the presence of the surrounding oxide support and an inert atmosphere well in excess of the Cu vapor pressure, any such melting is not expected to lead to any significant or irreversible effects on the nanoparticles. The particles then solidify once the light is turned off.

3.  Despite the above, even if we do assume that the melted layer drips off/evaporates, moves, merges with other particles etc., this will have only a slight change on the overall temperature distribution. Indeed, the absorption/heat source might become slightly distorted (its center shifted to a lower position within the layer), however, since clearly the final temperature distribution is determined primarily by the heat diffusion, the overall temperature distribution (and therefore the reaction rate) would change only slightly. This is confirmed in extensive numerical simulations we have been performing recently (not shown).


\end{document}